\documentclass[11pt,a4paper]{article}
\usepackage{amssymb}
\usepackage{jheppub}

\newcommand{\bea}{\begin{eqnarray}}
\newcommand{\eea}{\end{eqnarray}}
\newcommand{\bean}{\begin{eqnarray*}}
\newcommand{\eean}{\end{eqnarray*}}
\newcommand{\nn}{\nonumber\\}

\def\W #1{\widetilde{#1}}
\def\WH #1{\widehat{#1}}

\def\braket#1{\left\langle #1 \right\rangle}

\def\gb #1{ \left\langle #1 \right]}

\def\a{{\alpha}}

\def\b{{\beta}}

\def\la{\lambda}

\def\vev{\braket}

\def\bket#1{\left| #1\right]}
\def\bvev#1{\left[ #1 \right]}
\def\Spaa{\vev}
\def\Spbb{\bvev}
\def\Spab{\gb}

\def\Label#1{\label{#1}%
  \smash{\hbox to0pt{\raise1ex\hbox{\tiny[#1]}\hss}}}

\def\bse{\begin{subequations}}
\def\ese{\end{subequations}}
\def\eqref{\ref}
\def\Sl{\sum\limits}

\allowdisplaybreaks

\title{On tree amplitudes with gluons coupled to gravitons}

\author[a]{Yi-Xin Chen}
\author[a,c]{Yi-Jian Du}
\author[b,d]{Bo Feng}
\affiliation[a]{Zhejiang Institute of Modern Physics, Department of
Physics, Zhejiang University,\\
Hangzhou, 310027, P. R. China} \affiliation[b]{Center of
Mathematical Science, Zhejiang
University,\\
Hangzhou, 310027, P. R. China}
\affiliation[c]{Kavli Institute for Theoretical Physics China,Chinese Academy of Sciences,\\
Beijing 100190, P. R. China}
 \affiliation[d]{Key Laboratory of
Frontiers in Theoretical Physics, Institute of Theoretical Physics,
Chinese Academy of Sciences,\\Beijing 100190, P. R. China}
\abstract{In this paper, we study the tree amplitudes with gluons
coupled to gravitons. We first study the relations among the mixed
amplitudes. With BCFW on-shell recursion relation, we will show the
color-order reversed relation, $U(1)$-decoupling relation and KK
relation hold for tree amplitudes with gluons coupled to gravitons.
We then study the disk relation which expresses mixed amplitudes by
pure gluon amplitudes. More specifically we will prove the disk
relation for mixed amplitudes with gluons coupled to one graviton.
Using the disk relation and the properties of pure gluon amplitudes,
the color-order reversed relation, $U(1)$-decoupling relation and KK
relation for mixed amplitudes can also be proved. Finally, we give
some brief discussions on BCJ-like relation for mixed amplitudes. }

\begin{document}
%%%%%%%%%%%%%%%%%%%%%%%%%%%%%%%%%%%%%%%%%%%%%%%%%%%%%%%%%%%%%%%%%%
\maketitle

%%%%%%%%%%%%%%%%%
\section{Introduction}
%%%%%%%%%%%%%%%%%
Gauge field and gravity are two important objects in theoretical
physics. They contain all the fundamental interactions in nature. To
study the properties of gauge field and gravity, we should
investigate the perturbative scattering amplitudes in both theories.
The amplitude relations play an important role in both theories.

In gauge field theory, many relations among pure gluon tree partial
amplitudes have been found. They are color-order reversed relation,
$U(1)$-decoupling identity,  Kleiss-Kuijf(KK) relation
\cite{Kleiss:1988ne} and  Bern-Carrasco-Johansson(BCJ) relation
\cite{Bern:2008qj}:

Color-order reversed relation for pure gluon tree amplitudes is \bea
A(g_1,...,g_n)=(-1)^{n}A(g_n,...,g_1).~~~\Label{C-O-R-pure} \eea
$U(1)$-decoupling identity for pure gluon tree amplitudes is \bea
A(g_1,g_2,...,g_n)+A(g_1,g_3,g_2,...,g_n)+...
+A(g_1,g_3,...,g_n,g_2)=0.~~~\Label{U(1)-D-I-pure}
 \eea
KK relation for pure gluon tree amplitudes is \bea
A(g_1,\{\alpha\},g_l,\{\beta\})=(-1)^{n_{\beta}}\sum\limits_{\sigma\in
OP(\{\alpha\},\{\beta^T\})}A(g_1,\sigma,g_l).~~~\Label{KK-pure} \eea
Fundamental BCJ relation\footnote{Other BCJ relations can be derived
by fundamental BCJ relation combining with KK relation.} for pure
gluon tree amplitudes is \bea
&&s_{g_2g_3}A(g_1,g_3,g_2,...,g_n)+(s_{g_2g_3}+s_{g_2g_4})A(g_1,g_3,g_4,g_2,...,g_n)\nn
&+&...+(s_{g_2g_3}+s_{g_2g_4}+...+s_{g_2g_n})A(g_1,g_3,...,g_n,g_2)=0.\Label{BCJ-pure}
\eea The color-order reversed relation and $U(1)$-decoupling
identity can be regarded as two special cases of KK relation. To see
this we can choose $l=2$, $\{\alpha\}=\phi$ and
$\{\beta\}=\{g_3,...,g_n\}$ in (\ref{KK-pure}), we get color-order
reversed relation (\ref{C-O-R-pure}). When we choose $l=n-1$,
$\{\alpha\}=\{g_2,...,g_{n-2}\}$ and $\{\beta\}=\{g_n\}$, we get the
$U(1)$-decoupling identity (\ref{U(1)-D-I-pure}). Being different
from the other three relations, the BCJ relation have nontrivial
factors before the amplitudes.  The factors are constructed by
$s_{g_ig_j}$. With KK and BCJ relations, the pure gluon amplitudes
can be expressed by $(n-3)!$ independent amplitudes. Both KK
relation and BCJ relation can be understood by the  monodromy in
string theory \cite{BjerrumBohr:2009rd, Stieberger:2009hq}. The
field theory proof of these relations are given in
\cite{DelDuca:1999rs,Feng:2010my,Tye:2010kg}(See \cite{Jia:2010nz}
for $N=4$ SYM).

In general relativity, there is Kawai-Lewellen-Tye(KLT) relation
\cite{Kawai:1985xq} connects the gravity amplitudes to the gauge
field amplitudes. This relation was first found in string theory by
the study of closed string amplitudes on sphere. Taking the field
theory limit, KLT relation express the $m$-graviton amplitudes by
sum of products of two $m$-gluon amplitudes with appropriate
factors\cite{Berends:1988zp,Bern:1998ug}. KLT relation in field
theory has been proved in
\cite{BjerrumBohr:2010ta,BjerrumBohr:2010zb, BjerrumBohr:2010yc}(see
\cite{Feng:2010br} for $N=8$ SUGRA). It can also be used to treat
many amplitudes with gravitons coupled to matter
\cite{Bern:1999bx,BjerrumBohr:2004wh}. KLT relation is conjectured
to be hold at loop level in \cite{KLT-loop}. Recent researches
\cite{KLT-understanding} give a further understanding\footnote{Many
of them due to the BCJ relation.} of KLT relation.

Britto-Cachazo-Feng-Witten(BCFW) recursion relation
\cite{Britto:2004ap,Britto:2005fq} is an important recursion
relation which constructs on-shell tree amplitudes by on-shell tree
amplitudes with less external legs£º \bea
M_n=\Sl_{\mathcal{I},\mathcal{J},\mathbf{h}}\frac{M_{\mathcal{I}}(\WH
p_i,\WH P_{\mathcal{I},\mathcal{J}}^{\mathbf{h}})M_{\mathcal{J}}(\WH
p_j,-\WH
P_{\mathcal{I},\mathcal{J}}^{-\mathbf{h}})}{P_{\mathcal{I}\mathcal{J}}},
\Label{BCFW} \eea where the sum is over all possible distributions
of the external legs with shifted momentum $\WH p_i$ in
$\mathcal{I}$ and $\WH p_j$ in $\mathcal{J}$.
$z_{\mathcal{I}\mathcal{J}}$ indicates the splitting. Via
discussions on complex analysis, this relation exists in the
theories with $M(z\rightarrow\infty)=0$. Gauge field theory, general
relativity and the theory of gauge field coupled to gravity have
good behavior when $z\rightarrow \infty$ \cite{z-infinite}, thus in
all the three theories, one can use BCFW relation. Though in this
paper we only use the BCFW at tree-level in field theory, there are
works on BCFW recursion relation at loop-level
\cite{BCFW:loop-level} and in string theory\cite{BCFW:string}. The
discussion on BCFW recursion relation with nontrivial boundary
conditions are considered in \cite{BCFW:boundary}.

Since the on-shell three point amplitudes can be determined by the
symmetry of S-matrix \cite{S-matrix-program}, with BCFW recursion,
one can derive the on-shell tree amplitudes from on-shell three
point amplitudes. Using the methods in S-matrix program, one can
avoid the complicated Feynman diagrams. In fact, the three point
amplitudes and BCFW recursion relations in S-matrix program play a
crucial role in the proof of the amplitude relations in\footnote{In
\cite{DelDuca:1999rs}, KK relation is proved via new color
decomposition.} \cite{Feng:2010my,Tye:2010kg}, \cite{Jia:2010nz},
\cite{BjerrumBohr:2010ta,BjerrumBohr:2010zb, BjerrumBohr:2010yc} and
\cite{Feng:2010br}.

Though there are a lot of works on gauge field theory and general
relativity, the  theory with gauge field coupled to gravity is also
important. In this paper, we study the amplitudes with gluons
coupled to gravity. We will consider two categories of relations for
amplitudes with gluons coupled to gravitons:
\begin{itemize}
\item{(1)} The relations among the mixed amplitudes.
\end{itemize}
\begin{itemize}
\item{(2)}  The relation between the mixed amplitudes and pure gluon amplitudes.
\end{itemize}

In the first case, we will show the color-order reversed relation,
$U(1)$-decoupling identity and KK relation for pure gluon amplitudes
also hold for amplitudes with gluons coupled to gravitons. This is
because the gravitons do not effect the color structure of the gauge
field. However, the BCJ relation is more complicated. There are
nontrivial factors before the amplitudes. The formula of the BCJ
relation for pure gluon amplitudes do not hold for mixed amplitudes.

In the second case, there is disk relation for amplitudes with $n$
gluons coupled to $m$ gravitons which express the $(n,m)$ mixed tree
amplitudes by $n+2m$-point pure gluon tree amplitudes. This relation
was first considered in string theory
\cite{Chen:2009tr,Stieberger:2009hq} and have been shown to hold for
MHV amplitudes with only two negative helicity
gluons\cite{Chen:2010sr}. In \cite{Chen:2010sr}, a helicity
independent formula of disk relation for $(n,1)$ amplitudes is
conjectured. In this paper, we will use BCFW recursion relation and
the relations among pure gluon amplitudes to prove the relation for
$(n,1)$ amplitudes. Using the disk relation for amplitudes with $n$
gluons coupled to one graviton and the relations among pure gluon
amplitudes, we can give another proof of the color-order reversed
relation, $U(1)$-decoupling identity and KK relation.

The structure of this paper is as follows. In section
\ref{Sect-Gen-Dis}, we will discuss the mass dimension of the
amplitude. In section \ref{Sect-3-point}, we will give the three
point amplitudes for mixed amplitudes by S-matrix program. In
sections \ref{Sect-C-O-R}, \ref{Sect-U(1)-D-I} and \ref{Sect-KK} we
will prove the color-order reversed relation, $U(1)$-decoupling
identity and KK relation for mixed amplitudes by BCFW recursion
relation. We will prove the disk relation for mixed amplitudes with
only one graviton in section \ref{Sect-disk}. In the sections,
\ref{Sect-C-O-R-disk}, \ref{Sect-U(1)-D-I-disk} and
\ref{Sect-KK-disk}, we will give another approach to the color-order
reversed relation, $U(1)$-decoupling identity
 and KK relation for mixed amplitudes with only one graviton by disk relation.
At last, in \ref{Sect-remarks}, we will give some remarks on the BCJ
relation for mixed amplitudes and disk relation for mixed amplitudes
with more gravitons.
%%%%%%%%%%%%%%%%%
\section{Mass dimension of amplitude}\label{Sect-Gen-Dis}
%%%%%%%%%%%%%%%%%

As we have said, in this paper we will use the BCFW on-shell
recursion relation plus some general principles in quantum field
theory to discuss the tree-level amplitudes of gluons coupled to
gravitons. These general principles include Lorentz invariance,
gauge symmetry, dimensional analysis and analytic property etc. We
will discuss the implication of these general principles one by one,
but in this section we will focus on the analysis of mass dimension.

%In this section, we will give some general discussions on amplitudes with
%gluons coupled to gravitons.

{\bf The mass-dimension of pure gluon amplitude:} The mass dimension
of pure gluon tree-level scattering amplitude is
$(4-n)$\footnote{The mass dimension of spinor is ${1\over 2}$, thus
the Lorentz invariant contraction $\Spaa{1|2}$ has   the
mass-dimension  one.}. The reason for this result is following.
First the  mass dimension of gauge  field $A$ is one.  For the
calculation of scattering amplitudes,  we have replaced off-shell
field $A$ by on-shell polarization vector (the wave function) which
is dimension zero(for example ${\la\W\la\over \Spaa{q|\la}}$), thus
each time we add one external particle, we have reduced the mass
dimension by one. Secondly, the inner gluon propagator is given by
following replacement $AA\to {1\over p^2}$, so each gluon propagator
reduces the mass dimension by four. Using above two observations and
the starting point that each vertex is mass dimension four (so under
the $\int d^4 x$ will give mass dimension zero), we reach the
conclusion that the mass dimension of $n$ gluons is $(4-n)$.
Furthermore  the overall coupling constant of $n$ gluons is
$g^{n-2}$ where $g$ is the coupling constant of gauge theory.

{\bf The mass-dimension of graviton:} The mass dimension of graviton
field $h$ is again one. The big difference from gauge theory is that
the coupling constant $\kappa$ has mass dimension minus one.
Similarly to gauge theory, each time we add one external graviton
the total mass dimension is reduced by one which each inner
propagator of graviton reduces the mass dimension by four. Combining
these together we know immediately that the total mass dimension of
$n$ graviton amplitude is $(4-n)$ with the form $\kappa^{n-2} {\cal
M}$, thus the mass dimension of ${\cal M}$ is always $2$. This is
consistent with the KLT relation where we have ${\cal M}_n\sim A_n^2
s^{n-3}$, so the mass dimension by KLT relation is
$2(4-n)+2(n-3)=2$.

{\bf The mass dimension of gluon coupling to graviton:} Now we
consider the mixed case where $n$ gluons couple to  $m$ gravitons.
For this case, the mass dimension analysis is a little bit
complicated. To see this, let us start with three-point vertex. The
dimensional analysis, Lorentz invariance plus the weak coupling
limit tell us that the leading contribution must be $\kappa
\partial^2 A^2 h$. Using this to construct the effective four-point
vertex, we obtain $\kappa^2 \partial^2 A^4$ or  $\kappa^2 \partial^2
A^2 h^2$. Thus we see that even for $A^4$ vertex, there are two of
them: one from pure gauge theory with coupling $g^2$ and one by
contraction of gravitons with coupling $\kappa^2 s_{ij}$. In string
theory, we have $g\sim \kappa \ell$ (where $\ell$ is string scale)
and both terms are equivalent important. However, in field theory
these two coupling constants are independent to each other and we
can take different ordering. In this paper, we will take the limit
$g\gg \kappa$, thus the leading contribution of given external
particle configurations are these without inner graviton
propagators. In this paper, we will focus on these leading
contributions which can be used to compare with these obtained from
string theory by taking the field theory limit. The general form of
these leading terms are $g^{n-2}\kappa^{m} {\cal M}_{n;m}$ with mass
dimension $[{\cal M}_{n;m}]=4-n$. As we will show later, we have
${\cal M}_{n;m}\sim s^m A_{n+2m}$ and it is easy to check the mass
dimension is indeed right.

We want to emphasize one thing also. For our study of mapping, the
amplitude of pure gluon part is not general, i.e., the particles
corresponding to gravitons having special momenta. Thus what we have
is a special limit of these pure gluon amplitude. As we will see,
there is some nontrivial thing happening.

%%%%%%%%%%%%%%%%%
\section{The three point amplitudes}\label{Sect-3-point}
%%%%%%%%%%%%%%%%%

In this paper, we try to make the discussion as general as possible,
particularly we do not want to use the explicit Langrangian formula.
The only unfamiliar assumption we have made is the applicability of
BCFW on-shell recursion relation for the calculation of tree level
scattering amplitude. To be able to do recursive calculations, there
is a staring point, i.e., the three point amplitude. In this section
we will show that for three point amplitudes of given massless
particles with definite helicities, Lorentz symmetry, spin symmetry,
gauge symmetry plus the mass dimension  almost uniquely fixed them
to be (See the paper \cite{S-matrix-program}) a particular form. Let
us see how this is reached.

First, momentum conservation tells us that we should focus only on
the holomorphic part or anti-holomorphic part. From the Lorentz
symmetry and spin symmetry we have following differential equation
for three-point amplitude
\bea \left( \Spaa{\la_i|{\partial\over \partial \la_i}}+2
\mathbf{h}_i\right) M_3^{H}
(\Spaa{1|2},\Spaa{2|3},\Spaa{3|1})=0,~~~\forall
i~~~\Label{helicity-eq}\eea
where
\bea F_3=
\Spaa{1|2}^{d_3}\Spaa{2|3}^{d_1}\Spaa{3|1}^{d_2},~~~~d_i=\mathbf{h}_i-\mathbf{h}_{j}-\mathbf{h}_k,
~~~i,j,k=1,2,3,~~i\neq j\neq k\eea
is a particular solution of (\ref{helicity-eq}) with mass dimension
$\mathbf{h}_1+\mathbf{h}_2+\mathbf{h}_3$. Now we write $M_3= F_3
G_3$ with arbitrary function $G_3$, thus the equation for $G_3$ is
\bea \left( \Spaa{\la_i|{\partial\over \partial \la_i}}\right)
G_3^{H} (\Spaa{1|2},\Spaa{2|3},\Spaa{3|1})=0,~~~\forall
i~~~\Label{helicity-eq-G}\eea
which up to some delta-function case, has only {\sl constant
solution} with mass dimension $[M_3]-(h_1+h_2+h_3)$\footnote{Here we
have excluded the case where we can construct the Lorentz invariant
contraction other that $\Spaa{~|~}$. For example, in the
non-commutative field theory we can construct $k_\mu q_\nu
\theta^{\mu\nu}$ with antisymmetric constant $\theta$.}. This lead
us  to write down
\bea M_3= \kappa_H
\Spaa{1|2}^{d_3}\Spaa{2|3}^{d_1}\Spaa{3|1}^{d_2}+\kappa_A
\Spbb{1|2}^{-d_3}\Spbb{2|3}^{-d_1}\Spbb{3|1}^{-d_2},~~~\Label{M3-gen}
\eea
where $\Spaa{i|j}\equiv (\lambda_i)_a(\lambda_j)_b\epsilon^{ab}$ and
$\Spab{i|j}\equiv
(\W\lambda_i)_{\dot{a}}(\W\lambda_j)_{\dot{b}}\epsilon^{\dot{a}\dot{b}}$.
Applying (\ref{M3-gen}) to various situations we can obtain many
interesting results. For example, if three particles are gluons,
(\ref{M3-gen}) implies $A(1,2,3)=-A(3,2,1)$ for color-order
amplitude. Staring from this simple result plus the BCFW recursion
relation, general color-order reversed relation, the
$U(1)$-decoupling relation, the KK-relation as well as the new
discovered BCJ relation are all proved easily. Similarly, if three
particles are gravitons, (\ref{M3-gen}) implies
$M_3(1,2,3)=A_3(1,2,3)^2$ there $M_3$ is three-point amplitude of
gravitons. Using this result plus the BCFW recursion relation, the
KLT relation is proved recently. In our mixed case, there are also
some nonzero configurations, for example the one with two gluons and
one graviton ${A}_{2;1}$\footnote{In this paper, we use $M$ and $A$
to denote the total amplitudes and partial amplitudes ,respectively.
For amplitudes with $n$ gluons and $m$ graviton, the two amplitudes
are connected by color decomposition:
$M(g_1^{a_1},...,g_n^{a_n};H_1,...,H_m)=\Sl_{\sigma}Tr(T^{a_{\sigma(1)}}...T^{a_{\sigma(n)}})A(g_{\sigma(1)},...,g_{\sigma(n)};H_1,...,H_m)$
.}. For this one, we have ${ A}(g_1, g_2;H_1)={\cal A}(g_2,g_1;
H_1)$ which is trivially true. This relation will be used in our
late proof of some relations.

Having above general discussions, we will give more details for
different configurations.  Since results for pure  gluons or pure
gravitons are well known, we will focus only on the mixed case
${\cal A}_{n,m}$ with mass dimension $(4-n)$.

%%%%%%%%%%%%%%%%%%%%%%%%%%%%%%%
\subsection{One gluon plus two gravitons}
%%%%%%%%%%%%%%%%%%%%%%%%%%%%%%%

The simplest case is one gluon with two gravitons with mass
dimension three. Let us proceed it by dimension analysis. Using
parity symmetry we can focus on following three helicity
configurations.

{\bf Helicity $A(g_1^+; H_1^{++}, H_2^{++})$:} Since
$\sum\mathbf{h}_i=+5$ and $d_1=-3, d_2=d_3=-1$, from (\ref{M3-gen})
the possible nonzero contribution is the factor
$\Spbb{g_1|H_1}\Spbb{H_1|H_2}^{3} \Spbb{H_2|g_1}$, which has the
dimension five, times a factor which is helicity neutral and mass
dimension minus two. The only possibility  is to insert $s_{ij}$ at
the denominator, which is  singular. So we know this helicity will
have zero amplitude.

{\bf Helicity $A(g_1^+; H_1^{++}, H_2^{--})$:} Since
$\sum\mathbf{h}_i=+1$ and $d_1=1, d_2=3, d_3=-5$, the possible
nonzero contribution is the factor
$\Spbb{g_1|H_1}^{5}\Spbb{H_1|H_2}^{-1} \Spbb{H_2|g_1}^{-3}$, which
has the dimension one, times a factor which is helicity neutral and
mass dimension  two. Thus  we need to insert a factor $s_{ij}$ in
numerator, which will give zero result on-shell.

{\bf Helicity $A(g_1^-; H_1^{++}, H_2^{++})$:} Since
$\sum\mathbf{h}_i=+3$ and $d_1=-5, d_2=d_3=+1$, from (\ref{M3-gen})
the possible nonzero contribution is the factor
$\Spbb{g_1|H_1}^{-1}\Spbb{H_1|H_2}^{5} \Spbb{H_2|g_1}^{-1}$, which
has the dimension three, times a factor which is helicity neutral
and mass dimension zero. Thus this configuration could be nonzero.
Thus this example shows that just dimension analysis plus Lorentz
symmetry can not exclude its existence, the gauge symmetry must be
used. The reason that we must have at least two gluons to be gauge
invariant is because non-abelian gluons carry color charges, so the
charge neutral condition requires this.

%%%%%%%%%%%%%%%%%%%%%%%%%%%%%%%
\subsection{Two gluons plus one graviton}
%%%%%%%%%%%%%%%%%%%%%%%%%%%%%%%

By parity symmetry we can fix graviton to be positive, thus we have
following three helicity configurations with mass dimension two to
consider.

{\bf Helicity $A(g_1^+, g_2^+ ;H_1^{++})$:} By general argument, it
should be given by factor $\Spbb{g_1|g_2}^0 \Spbb{g_2|H_1}^2
\Spbb{g_1|H_1}^2$ with  mass dimension four times another helicity
neutral factor with mass dimension minus two. The only way is to
insert a factor $s_{ij}$ in denominator which will be singular. Thus
the amplitude should be zero.

{\bf Helicity $A(g_1^+, g_2^+; H_1^{--})$:} This is the tricky case
since $\sum \mathbf{h}_i=0$. The holomorphic part has factor
$\Spaa{g_1|g_2}^{-4}\Spaa{g_1|H_1}^{2} \Spaa{g_2|H_1}^2$ while
anti-holomorphic part has factor
$\Spbb{g_1|g_2}^{4}\Spbb{g_1|H_1}^{-2} \Spbb{g_2|H_1}^{-2}$. Both
have the mass dimension zero while we need mass dimension two, so
the only way is to insert a $s_{ij}$ factor. Thus no matter which
part (holomorphic or anti-holomorphic) we take,  the insertion of
$s_{ij}$ will give zero result. This is also consistent with our
claim  $A(g_1^+, g_2^+; H_1^{--})\sim s_{g_1g_2} A(g_1^+, g_2^+,
h_1^-, \W h_1^-)$ (See the disk relation given in
\cite{Chen:2010sr}) as we will discuss late.

{\bf Helicity $A(g_1^+, g_2^-; H_1^{++})$:} By general argument  it
should be factor $\Spbb{g_1|g_2}^{-2} \Spbb{g_2|H_1}^0
\Spbb{g_1|H_1}^4$ with mass dimension two times another helicity
neutral factor with dimension zero. Thus this configuration can have
nonzero result and in fact, it does.

Let us discuss this helicity configuration more carefully. By the
mapping relation we will present late, we will have $A(g_1^+, g_2^-;
H_1^{++})= s_{g_1g_3} A_4(g_1^+, h_1^+, g_2^-, \W h_1^+)$ where the
momentum $p_{h_1}=p_{\W h_1}={1\over 2} p_{H_1}$. Naively since
$A_4(g_1^+, h_1^+, g_2^-, \W h_1^+)$ has only one negative helicity
and  is zero. However, the whole conclusion is very tricky and let
us do some general analysis based on symmetry and dimension.

%because we are focus only on very special momentum configuration
%where $p_{h_1}=p_{\W h_1}={1\over 2} p_{H_1}$.

By spin symmetry and mass dimension we get following general
expression
\bea & &  s_{g_1h_1}^{a_{13}} s_{g_1g_2}^{b_{12}}
s_{g_2h_1}^{b_{23}} s_{g_2\W h_1}^{a_{24}}
[\Spaa{g_1|g_2}\Spaa{h_1|\W h_1}\Spbb{g_1|h_1}\Spbb{g_2|\W
h_1}]^{a_{34}} [ \Spaa{g_2|h_1}\Spaa{g_1|\W
h_1}\Spbb{g_1|h_1}\Spbb{g_2|\W h_1}]^{a_{14}} \nn &\times &\left(
{\Spbb{g_1|\W h_1}\over \Spaa{g_2|h_1}\Spbb{g_2|\W
h_1}\Spbb{g_1|h_1}}\right)^{b_{14}} \left( {\Spbb{h_1|\W h_1}\over
\Spaa{g_1|g_2} \Spbb{g_1|h_1}\Spbb{g_2|\W h_1}}\right)^{b_{23}}
[\Spaa{g_1|g_2}\Spaa{g_2|h_1}\Spbb{g_1|h_1}^2\Spbb{g_2|\W h_1}],\eea
where we have used the on-shell condition for four gluons (i.e.,
$p_i^2=0$), but not the momentum conservation condition as well as
the color ordering information. Parameters like $b,a$ are free
integers. Using the momentum conservation, we can write \bea
\Spaa{g_2|h_1}\Spaa{g_1|\W h_1}\Spbb{g_1|h_1}\Spbb{g_2|\W
h_1}=\Spaa{g_2|\W h_1}\Spaa{g_1|\W h_1}\Spbb{g_1|\W h_1}\Spbb{g_2|\W
h_1}, \eea
 and similarly another factor. Furthermore, ${\Spbb{g_1|\W h_1}\over
\Spaa{g_2|h_1}\Spbb{g_2|\W h_1}\Spbb{g_1|h_1}}={\Spbb{g_1|\W
h_1}\over \Spaa{g_1|h_1}\Spbb{g_1|\W h_1}\Spbb{g_1|h_1}}$ and
similarly another factor. Having done all these we will have
following expression
\bea A(g_1^+, h_1^+, g_2^-,  \W h_1^+) \sim s_{g_1h_1}^{a_{13}}
s_{g_1g_2}^{b_{12}} s_{g_1\W
h_1}^{a_{14}}[\Spaa{g_2|g_1}\Spbb{g_1|h_1}\Spbb{g_1|\W h_1}]^2\sim
s_{g_1h_1}^{a_{13}} s_{g_1g_2}^{b_{12}} s_{g_1\W h_1}^{a_{14}}\left(
{\Spbb{g_1|h_1}\Spbb{g_1|\W h_1}\over \Spbb{g_1|g_2}}\right)^2,\eea
where we have used the $s_{g_1g_2}=s_{h_1\W h_1}$ etc. The
color-ordering information tell us that we should have pole
$s_{g_1h_1}$ and $s_{g_1\W h_1}$ with power at most one in
denominator. Finally mass dimension zero tells us that the solution
should  be
\bea A(g_1^+, h_1^+, g_2^- , \W h_1^+) \sim {s_{g_1g_2}\over
s_{g_1\W h_1} s_{g_1h_1}}\left( {\Spbb{g_1|h_1}\Spbb{g_1|\W
h_1}\over \Spbb{g_1|g_2}}\right)^2.~~~~\Label{A4-gen}\eea
Formula (\ref{A4-gen}) is well-defined if momenta of four particles
are general distributed. However, for our consideration,  momentum
configuration is very special $p_{h_1}=p_{\W h_1}={1\over 2}
p_{H_1}$. At this special point, $A(g_1^+, h_1^+, g_2^- , \W h_1^+)$
is singular and we should put the whole gluon scattering amplitude
to zero.

For $A_{2;1}$, thing is different since we will have
\bea  A(g_1^+, g_2^-; H_1^{++})\sim s_{g_1h_1} A(g_1^+, h_1^+,
g_2^-,  \W h_1^+) \sim {s_{g_1g_2}\over s_{g_1\W h_1} }\left(
{\Spbb{g_1|h_1}\Spbb{g_1|\W h_1}\over \Spbb{g_1|g_2}}\right)^2.\eea
After using momentum conservation $s_{g_1g_2}=-s_{g_1h_1}-s_{g_1\W
h_1}$ and finally the degenerate limit $p_{h_1}=p_{\W
h_1}$\footnote{ It is worth to notice that we have used the momentum
conservation before taking the $p_{h_1}=p_{\W h_1}$ limit.}, we
obtain
\bea A(g_1^+, g_2^-; H_1^{++})\sim {\Spbb{g_1|h_1}^4\over
\Spbb{g_1|g_2}^2}\eea
which is well defined.

We can have another understanding by  using the BCFW recursion
relation. Using the BCJ relation,
 the mapping  can be write to another
 form $A(g_1^+, g_2^-; H_1^{++})\sim
 s_{g_1g_2} A_4(g_1^+,
g_2^-, h_1^+, \W h_1^+)$.
%Naively it is zero for both factor
%$s_{g_1g_2}$ and $A_4(g_1,g_2,h_1,\W h_1)$ if it is general
%four-point amplitude. However, as we have emphasized, we are at the
%special  phase space point where $p_{h_1}=p_{\W h_1}={-1\over 2}
%(p_{g_1}+p_{g_2})$. Thus there is potential infinity from $A_4$.
Let us calculate the right hand side by taking the BCFW
$\Spab{g_1|g_2}$-deformation, i.e, $\W\la_{g_2}\to
\W\la_{g_2}+z\W\la_{g_1}$ and $\la_{g_1}\to \la_{g_1}-z\la_{g_2}$.
The $A_4$ part  is given by
\bean {\Spbb{h_1|P}^3\over \Spbb{P|\WH g_2}\Spbb{\WH g_2|P}}{1\over
s_{g_1g_2}} {\Spbb{\W h_1|g_1}^3\over \Spbb{g_1|P}\Spbb{P|\W h_1}}
\eean
with $z={-\Spbb{h_1|g_2}\over \Spbb{h_1|g_1}}$ and $\bket{\WH
g_2}={\Spbb{g_1|g_2}\over \Spbb{g_1|h_1}}\bket{h_1}$,
$\bket{P}=\bket{h_1}$. Putting it back and  using $\bket{\W
h_1}=\bket{h_1}$ in our special momentum configuration we have
${\Spbb{g_1|h_1}^4\over \Spbb{g_1|g_2}^2} {1\over s_{g_1g_2}}$.
Multiplying the factor back we get immediately
\bea A(g_1^+, g_2^-; H_1^{++})\sim s_{g_1g_2} A_4(g_1^+, g_2^-,
h_1^+, \W h_1^+) \sim {\Spbb{g_1|h_1}^4\over \Spbb{g_1|g_2}^2}.\eea

One important consequence is that
\bea A(g_1^-, g_2^+; H_1^{++})\sim s_{g_1g_2} A_4(g_1^-, g_2^+,
h_1^+, \W h_1^+) \sim {\Spbb{g_2|h_1}^4\over \Spbb{g_1|g_2}^2}\eea
thus we have
\bea A(g_1^+, g_2^-; H_1^{++})= A_{2;1}(g_2^-, g_1^+; H_1^{++}).\eea
Although it is obviously true because ${\rm Tr}(T_1 T_2)= {\rm
Tr}(T_2 T_1)$, it will play important role in proofs given in late
section.

Finally we want to emphasize again that the mapping we have used in
this part is defined only at special momentum configurations where
$p_{h_1}=p_{\W h_1}$.

%%%%%%%%%%%%%%%%%%%%%%%%%%%%%%%
\section{The color-order reversed relation}\label{Sect-C-O-R}
%%%%%%%%%%%%%%%%%%%%%%%%%%%%%%%

Having done some general discussions, we will move to the proof of
some identities of tree-level scattering amplitudes of color-ordered
gluons coupling to gravitons at the leading order, i.e., we have
neglected contributions with graviton propagators. The proof are
similarly with in the pure gluon case\cite{Feng:2010my}. However,
the boundary case are more complicated since there are gravitons in
the amplitudes.

The first identity we will prove is the
 color-order reversed relation
\bea
A(g_1,...,g_n;H_1,...,H_m)=(-1)^{n}A(g_n,...,g_1;H_1,...,H_m),~~~n\geq
2, m\geq 1.~~~\Label{C-O-R-mixed} \eea
As we have remarked before, we have excluded the case $n=1$. Since
there are two integers $(n,m)$, the induction proof will go along
both directions, especially there are two boundary cases: one  is
$m=1$ with arbitrary $n\geq 2$ and another one,  $n=2$ with
arbitrary $m\geq 1$. To prove this inductively, we define $N=n+m$
and assume (\ref{C-O-R-mixed}) is true for any $3\leq \W N\leq N$,
then we prove (\ref{C-O-R-mixed}) for $N+1$. Furthermore, the
starting point $n=2,m=1$ for $N=3$ is obviously true as discussed in
previous section.

The organization of this section is following. First we give a
simple example. Secondly we deal with boundary case $m=1$ and
thirdly we deal with the boundary case $n=2$ and finally we give a
general proof for arbitrary $(n,m)$. Although for our general proof,
we do not need to prove two special cases, we present them here to
demonstrate our idea more clearly.

%%%%%%%%%%%%%%%%%%%%%%%%%%
\subsection{An example}
%%%%%%%%%%%%%%%%%%%%%%%%%%
The simplest example of color-order reversed relation for mixed
amplitudes is
\bea A(g_1,g_2,g_3;H_1)=(-1)A(g_3,g_2,g_1;H_1). \eea
To see this, we could expand the amplitude $A(g_1,g_2,g_3;H_1)$ by
BCFW recursion \cite{Britto:2004ap,Britto:2005fq} relation
as\footnote{In this paper we usually express the BCFW expansion
(\ref{BCFW}) as
$M_n=\Sl_{\mathcal{I},\mathcal{J},\mathbf{h}}M_{\mathcal{I}}(\WH
p_i,\WH P_{\mathcal{I},\mathcal{J}}^{\mathbf{h}}|-\WH
P_{\mathcal{I},\mathcal{J}}^{-\mathbf{h}},\WH
p_j)=M_{\mathcal{I}}(\WH p_i|\WH p_j)$ for short. We use
$A(H_1,...,H_i;\WH g_1,...,g_j|g_{j+1},...,\WH
g_{n};H_{i+1},...,H_m)$ to denote $\frac{A(\WH g_1,...,g_j,\WH
P;H_1,...,H_i)A(-\WH P,g_{j+1},...,\WH
g_{n};H_{i+1},...,H_m)}{P^2}$.}
\bean A(g_1,g_2,g_3;H_1)&=&A(\WH g_1,g_2,\WH P|-\WH P,\WH
g_3;H_1)+A(H_1;\WH g_1,\WH Q|-\WH Q,g_2,\WH g_3)
\\&=&(-1)^2A(H_1;\WH g_3,-\WH P|\WH P,g_2,\WH g_1)(-1)^2+(-1)^3A(\WH g_3,g_2,-\WH Q |\WH Q,\WH
g_1;H_1) (-1)^2
\\&=&(-1)^3A(g_3,g_2,g_1;H_1),
\eean
where the color-order reversed relation for pure gluon amplitudes as
well as the mixed amplitudes with only two gluons are used. Thus we
get the color-order reversed relation for amplitudes with three
gluons coupled to one graviton.

%%%%%%%%%%%%%%%%%%%%%%
\subsection{The boundary case with $m=1$}
%%%%%%%%%%%%%%%%%%%%%%

Now let us consider the boundary case with $m=1$ and arbitrary $n$.
The starting point $m=1,n=2$ is true by our general discussions.
Using BCFW expansion, we can write
\bean A(g_1,...,g_n;H_1)&=&A(\WH g_1,[g_2,...,g_{n-1}],\WH
g_n;H_1)+A(H_1;\WH g_1,[g_2,...,g_{n-1}],\WH g_n)
\\&=&(-1)^{n+2}A(H_1;\WH g_n,[g_{n-1},...,g_1],\WH g_1)+(-1)^{n+2}A(\WH g_n,[g_{n-1},...,g_1],\WH g_1;H_1)
\\&=&(-1)^nA(g_n,...,g_1;H_1),
\eean
where we have used $[g_2,...,g_{n-1}]$ to denote the sum over all
possible ordered splitting allowed by BCFW expansion. At the second
line we have used the color-order reversed relation for pure gluons
and for $m=1$ with less than $n$ gluons. At the third line we
recombine all component to give the amplitude $A(g_n,...,g_1;H_1)$.

%%%%%%%%%%%%%%%%%%%%%%
\subsection{The boundary case with $n=2$}
%%%%%%%%%%%%%%%%%%%%%%

Now let us consider the boundary case with $n=2$ and arbitrary
$m\geq 1$. The starting point $m=1,n=2$ is true by our general
discussions. Also that this particular case is true for general $m$
is also obviously because with only two elements, ${\rm Tr}(T_1
T_2)={\rm Tr}(T_2 T_1)$. We can also give it a proof by using BCFW
expansion as following. For a given distribution of $\{H\}$, e.g.
$\{H_1,...,H_i\}$ in the left set and $\{H_{i+1},...,H_m\}$ in the
right set, we have \bea A(g_1,g_2;H_1,...,H_m) &=&A(H_1,...,H_i;\WH
g_1|\WH g_2;H_{i+1},...,H_m)\nn &=&A(H_{i+1},...,H_m;\WH g_2|\WH
g_1;H_1,...,H_i).
 \eea
We should notice that in $A(g_2,g_1;H_1,...,H_m)$, there is a term
 $(H_{i+1},...,H_m;\WH g_2|\WH
g_1;H_1,...,H_i)$ in BCFW decomposition. Thus, after summing over
all the possible distributions of gravitons, we get the color-order
reversed relation for mixed amplitudes with only two gluons \bea
A(g_1,g_2;H_1,...,H_m)=A(g_2,g_1;H_1,...,H_m). \eea
%%%%%%%%%%%%%%%%%%%%%%%%%%
\subsection{Color-order reversed relation for amplitudes with gluons coupled to gravitons}
%%%%%%%%%%%%%%%%%%%%%%%%%%
Now we prove the general color-order reversed relation for
amplitudes with $n$ gluons and $m$ gravitons. The amplitude can be
given by BCFW recursion relation as a sum over all the possible
ordered splitting of the gluons and all possible distributions of
the gravitons. For a given distribution with $\{H_1,...,H_i\}$,
$\{g_2,...,g_k\}$ in the left set and $\{H_{i+1},...,H_{m}\}$,
$\{g_{k+1},...,g_n\}$ in the right set,  we should have
 \bea
 & &A(H_1,...,H_i;\WH g_1,g_2,...,g_k|g_{k+1},...,g_{n-1},\WH g_n;H_{i+1},...,H_m)\nn
 &=&(-1)^{n-k+1}A(H_{i+1},...,H_m;\WH g_n, g_{n-1},...,g_{k+1}|g_k,...,\WH
 g_1;H_1,...,H_i)(-1)^{k+1}\nn
 &=&(-1)^nA(H_{i+1},...,H_m;\WH g_n, g_{n-1},...,g_{k+1}|g_k,...,\WH
 g_1;H_1,...,H_i).
 \eea
 where we have used the color-order reversed relation for $N'<N$
 amplitudes. Thus this terms just correspond to a term in the BCFW decomposition
of $A(g_n, g_{n-1},...,
 g_1;H_1,...,H_m)$. After summing over all terms and using the color-order reversed relation for
 $N'<N$ amplitudes, we get the
 color-order reversed relation (\ref{C-O-R-mixed}).

%%%%%%%%%%%%%%%%%%%%%%%%%%%%%%%
\section{The $U(1)$-decoupling identity}\label{Sect-U(1)-D-I}
%%%%%%%%%%%%%%%%%%%%%%%%%%%%%%%
The second identity we try to prove is the $U(1)$-decoupling
identity for tree amplitudes with gluons coupled to gravitons. It is
given by
\bea
&&A(g_1,g_2,...,g_n;H_1,...,H_m)+A(g_1,g_3,g_2,...,g_n;H_1,...,H_m)+...\nn
&+&A(g_1,g_3,...,g_n,g_2;H_1,...,H_m)=0,~~~n\geq 3,m\geq
1~.~~~\Label{U(1)-D-I-mixed}
 \eea
 where we have moved $g_2$ to right one by one while keep the
 relative ordering of $(g_1,g_3,g_4,...,g_n)$. Similarly to previous
 section, our proof will go along $N=n+m$ inductively. The starting
 point is $n=3,m=1$ which we will prove in following.

%%%%%%%%%%%%%%%%%%%%%%%%%%
\subsection{An example}
%%%%%%%%%%%%%%%%%%%%%%%%%%

Let us start again with simple example where $n=3,m=1$. Using BCFW,
we can expand the three-gluon, one-graviton amplitudes as
\bea
 A(g_1,g_2,g_3;H_1)&=&A(H_1;\WH g_1,\WH P|-\WH P,\WH g_2,g_3)+A(g_3,\WH g_1,\WH Q|-\WH Q,\WH g_2;H_1),\nn
A(g_1,g_3,g_2;H_1)&=&A(H_1;\WH g_1,\WH P|-\WH P,g_3,\WH g_2)+A(\WH
g_1,g_3,\WH Q|-\WH Q,\WH g_2;H_1). \eea
Using the $U(1)$-decoupling identity for pure gluon $
A(g_1,g_2,g_3)+A(g_1,g_3,g_2)=0$, we get immediately
\bea A(g_1,g_2,g_3;H_1)+A(g_1,g_3,g_2;H_1)=0.
 \eea
 For the special case, we see the $U(1)$-decoupling identity for $A_{3,1}$ amplitudes
 is nothing, but   the color-order reversed
relation for $A_{3,1}$ amplitudes.

%%%%%%%%%%%%%%%%%%%%%%%%%%
\subsection{$U(1)$-decoupling identity for amplitudes with gluons coupled to gravitons}
%%%%%%%%%%%%%%%%%%%%%%%%%%
Now let us turn to the general $U(1)$-decoupling identity for
amplitudes with $n$ gluons coupled to $m$ gravitons. For the
amplitude $A(g_1,...,g_k,g_2,g_{k+1},...,g_n;H_1,...,H_m)$, we
should sum over all the possible distributions of the gravitons and
gluons. For a certain distribution of gravity, the sum over all
possible distribution of gluons is
 \bean&& A(H_1,...,H_i;\WH g_1,[g_3,...,g_k],\WH g_2,g_{k+1},...g_n;H_{i+1},...,H_m)\nn
 &+&A(H_1,...,H_i;g_n,\WH g_1,[g_3,...,g_k],\WH
 g_2,g_{k+1},...,g_{n-1};H_{i+1},...,H_m)\nn
 &+&...\nn
 &+&A(H_1,...,H_i;g_{k+1},...,g_n,\WH g_1,[g_3,...,g_k],\WH
 g_2;H_{i+1},...,H_m).
  \eean
For a given pole, the sum over all the possible positions of $g_2$
becomes the sum over all the possible positions of $\WH g_2$ in the
right set. Since we assumed there are $U(1)$-decoupling identities
for amplitudes with $N'<N$, we can use $U(1)$-decoupling identity in
the right set. For example \bea \Sl_{k=3}^nA(H_1,...,H_i;\WH
g_1,g_3|g_4,...,g_k,\WH g_2,g_{k+1},...,g_n;H_{i+1},...,H_{m})=0.
\eea After summing over all the possible poles, we get the
$U(1)$-decoupling identity.

We should discuss the special case with only $\WH g_2$ in the right
set when there are gravitons in the right set. In this case, there
are only two gluons in the right set, thus we cannot use the
$U(1)$-decoupling identity in the right set. For example \bea
A(H_1,...,H_i;g_{k+1},...,g_n,\WH g_1,g_3,...,g_k,\WH P|-\WH P,\WH
g_2;H_{i+1},...,H_m) \eea The sum over all the positions of $g_2$
becomes the sum over all the positions of $\WH P$ in the left set.
Thus we can use the $U(1)$-decoupling identity in the left set, the
sum of these terms also gives zero.

Considering all the contributions of poles, we get the
$U(1)$-decoupling identity (\ref{U(1)-D-I-mixed}).
%%%%%%%%%%%%%%%%%%%%%%%%%%%%%%%
\section{The KK relation}\label{Sect-KK}
%%%%%%%%%%%%%%%%%%%%%%%%%%%%%%%

The third identity we want to prove is the generalized KK relation
for amplitudes with gluons coupled to gravitons, which is given by
\bea
A(g_1,\{\alpha\},g_l,\{\beta\};H_1,...,H_m)=(-1)^{n_{\beta}}\sum\limits_{\sigma\in
OP(\{\alpha\},\{\beta^T\})}A(g_1,\sigma,g_l;H_1,...,H_m)~~~n\geq
3,m\geq 1.~~~\Label{KK-mixed} \eea
where the sum is over the ordered permutations of set $(\a,\b^T)$
where the relative ordering of elements of set $\a$ is kept (so as
the  set $\b^T$, where $T$ means the reversed ordering). The $n_\b$
is the number of set $\b$. If the set $\a$ is empty, it is the
color-order reversed relation. If the set $\a$ has only one element,
it is the $U(1)$-decoupling identity. Thus the KK-relation is more
general.

%%%%%%%%%%%%%%%%%%%%%%%%%%
\subsection{An example}
%%%%%%%%%%%%%%%%%%%%%%%%%%
The simplest KK relation for mixed amplitudes is $n=3,m=1$ which is
nothing, but the $U(1)$-decoupling identity we have proved.  Here we
give a nontrivial KK relation for amplitudes with six gluons coupled
to one graviton as an example. Using BCFW recursion relation, we
have \bea &&A(g_1,\{g_2,g_3\},g_4,\{g_5,g_6\};H_1)\nn &=&A(\WH
g_1,\{g_2\}|\{g_3\},\WH g_4,\{g_5,g_6\};H_1)+A(\{g_6\},\WH
g_1,\{g_2\}|\{g_3\},\WH g_4,\{g_5\};H_1)\nn &+&A(\{g_5,g_6\},\WH
g_1,\{g_2\}|\{g_3\},\WH g_4;H_1)+A(\WH g_1,\{g_2,g_3\}|\WH
g_4,\{g_5,g_6\};H_1)\nn &+&A(\{g_6\},\WH g_1,\{g_2,g_3\}|\WH
g_4,\{g_5\};H_1)+A(\{g_5,g_6\},\WH g_1,\{g_2,g_3\}|\WH g_4;H_1)\nn
&+&A(\{g_6\},\WH g_1,|\{g_2,g_3\},\WH
g_4,\{g_5\};H_1)+A(\{g_5,g_6\},\WH g_1,|\{g_2,g_3\},\WH g_4;H_1)\nn
&+&\text{terms with $H_1$ in the left set}. \eea Here, we choose the
momenta of $g_1$ and $g_4$ as shifted momenta. We can use KK
relation for amplitudes with less gluons in the left and the right
sets\footnote{In this example, the KK relations with less gluons are
just color-order reversed relation, $U(1)$-decoupling identity and
their combination.}. By induction, we get \bea
&&A(g_1,\{g_2,g_3\},g_4,\{g_5,g_6\};H_1)\nn &=&[A(\WH
g_1,\{g_2\}|\{g_3,g_6,g_5\},\WH g_4;H_1)+A(\WH
g_1,\{g_2\}|\{g_6,g_3,g_5\},\WH g_4;H_1)+A(\WH
g_1,\{g_2\}|\{g_6,g_5,g_3\},\WH g_4;H_1)]\nn &+&[A(\WH
g_1,\{g_2,g_6\}|\{g_3,g_5\},\WH g_4;H_1)+A(\WH
g_1,\{g_2,g_6\}|\{g_5,g_3\},\WH g_4;H_1)+\WH
g_1,\{g_6,g_2\}|\{g_3,g_5\},\WH g_4;H_1)\nn &+&A(\WH
g_1,\{g_6,g_2\}|\{g_5,g_3\},\WH g_4;H_1)]+[A(\WH
g_1,\{g_6,g_5,g_2\}|\{g_3\},\WH g_4;H_1)+A(\WH
g_1,\{g_6,g_2,g_5\}|\{g_3\},\WH g_4;H_1)\nn &+&A(\WH
g_1,\{g_2,g_6,g_5\}|\{g_3\},\WH g_4;H_1)]+A(\WH
g_1,\{g_2,g_3\}|\{g_6,g_5\},\WH g_4;H_1)+[A(\WH
g_1,\{g_2,g_3,g_6\}|\{g_5\},\WH g_4;H_1)\nn &+&A(\WH
g_1,\{g_2,g_6,g_3\}|\{g_5\},\WH g_4;H_1)+A(\WH
g_1,\{g_6,g_2,g_3\}|\{g_5\},\WH g_4;H_1)]+[A(\WH
g_1,\{g_2,g_3,g_6,g_5\}|\WH g_4;H_1)\nn &+&A(\WH
g_1,\{g_2,g_6,g_3,g_5\}|\WH g_4;H_1)+A(\WH
g_1,\{g_2,g_6,g_5,g_3\}|\WH g_4;H_1)+A(\WH
g_1,\{g_6,g_2,g_3,g_5\}|\WH g_4;H_1)\nn &+&A(\WH
g_1,\{g_6,g_2,g_5,g_3\}|\WH g_4;H_1)+A(\WH
g_1,\{g_6,g_5,g_2,g_3\}|\WH g_4;H_1)]+[A(\WH
g_1,\{g_6\}|\{g_2,g_3,g_5\},\WH g_4;H_1)\nn &+&A(\WH
g_1,\{g_6\}|\{g_2,g_5,g_3\},\WH g_4;H_1)+A(\WH
g_1,\{g_6\}|\{g_5,g_2,g_3\},\WH g_4;H_1)]++A(\WH
g_1,\{g_6,g_5\}|\{g_2,g_3\},\WH g_4;H_1)\nn &+&\text{terms with
$H_1$ in the left set}. \eea The equation above can be reexpressed
as \bea &&A(g_1,\{g_2,g_3\},g_4,\{g_5,g_6\};H_1)=A(\WH
g_1,[g_2,g_3,g_6,g_5],\WH g_4;H_1)+A(\WH g_1,[g_2,g_6,g_3,g_5],\WH
g_4;H_1)\nn &+&A(\WH g_1,[g_2,g_6,g_5,g_3],\WH g_4;H_1)+A(\WH
g_1,[g_6,g_2,g_3,g_5],\WH g_4;H_1)+A(\WH g_1,[g_6,g_2,g_5,g_3],\WH
g_4;H_1)\nn &+&A(\WH g_1,[g_6,g_5,g_2,g_3],\WH g_4;H_1)+\text{terms
with $H_1$ in the left set}. \eea The terms with $H_1$ in the left
set have similar expressions. Thus we get the expression of KK
relation \bea &&A(g_1,\{g_2,g_3\},g_4,\{g_5,g_6\};H_1)=A(\WH
g_1,g_2,g_3,g_6,g_5,\WH g_4;H_1)+A(\WH g_1,g_2,g_6,g_3,g_5,\WH
g_4;H_1)\nn &+&A(\WH g_1,g_2,g_6,g_5,g_3,\WH g_4;H_1)+A(\WH
g_1,g_6,g_2,g_3,g_5,\WH g_4;H_1)+A(\WH g_1,g_6,g_2,g_5,g_3,\WH
g_4;H_1)\nn &+&A(\WH g_1,g_6,g_5,g_2,g_3,\WH g_4;H_1). \eea

%%%%%%%%%%%%%%%%%%%%%%%%%%
\subsection{KK relation for amplitudes with gluons coupled to gravitons}
%%%%%%%%%%%%%%%%%%%%%%%%%%
Now we turn to the general discussion on KK relation for
$A(g_1,\{g_2,...,g_{l-1}\},g_l,\{g_{l+1},...,g_{n}\};H_1,...,H_m)$.
To construct this amplitude by BCFW recursion relation, we should
sum over all the possible poles, i.e. we should sum over all the
possible distributions of gravitons and gluons. For a given
distribution of gravitons and gluon e.g. we consider
$A(H_1,...,H_i;\{\beta_L\},\WH g_1, \{\alpha_L\}|\{\alpha_R\},\WH
g_l, \{\alpha_R\};H_{i+1},...,H_m)$. In this case, we can use  the
KK relation in both size
 \bea& & A(H_1,...,H_i;\{\beta_L\},\WH g_1,
\{\alpha_L\}|\{\alpha_R\},\WH g_l, \{\alpha_R\};H_1,...,H_m)\nn
&=&\Sl_{\sigma_L\in OP(\{\alpha_L\},\{\beta_L\}^T)}\Sl_{\sigma_R\in
OP(\{\alpha_R\},\{\beta_R\}^T)}(-1)^{n_{\beta_L}}A(H_1,...,H_i;\WH
g_1,\sigma_L|\sigma_R,\WH g_l;H_{i+1},...,H_m)(-1)^{n_{\beta_R}}.
 \eea
Summing over all the possible distributions of gluons, using the KK
relation in each term for both size, we get
\bea\Sl_{\text{distributions of }\{\alpha\} \text{and}
\{\beta\}}\Sl_{\sigma_L\in
OP(\{\alpha_L\},\{\beta_L\}^T)}\Sl_{\sigma_R\in
OP(\{\alpha_R\},\{\beta_R\}^T)}(-1)^{n_{\beta}}A(H_1,...,H_i;\WH
g_1,\sigma_L|\sigma_R,\WH g_l;H_{i+1},...,H_m)\label{KK-mixed1}
 \eea
This just gives \bea \Sl_{\sigma\in
OP(\{\alpha\},\{\beta^T\}}A(H_1,...,H_i;\WH g_1,[\sigma],\WH
g_n;H_{i+1},...,H_m).\label{KK-mixed2} \eea To see this, we should
count the number of terms in each expression as in
\cite{Feng:2010my}. In (\ref{KK-mixed1}), there are \bea
\Sl_{i=0}^{n_{\alpha}}\Sl_{j=0}^{n_{\beta}}C_{n_{\beta}-j+i}^{i}C_{n_{\alpha}-i+j}^{j}
=\Sl_{i=0}^{n_{\alpha}}\Sl_{j=0}^{n_{\beta}}\frac{(n_{\beta}-j+i)!}{i!(n_{\beta}-j)!}\frac{(n_{\alpha}-i+j)!}{j!(n_{\alpha}-i)!}.
\eea In (\ref{KK-mixed2}), there are
$\frac{(n_{\alpha}+n_{\beta})!}{n_{\alpha}!n_{\beta}!}(n_{\alpha}+n_{\beta}+1)$
terms, thus these to expression give the same number of terms. There
are two special cases when the gravitons are all in left or right
set. We take all gravitons in the left set as an example: \bea
\Sl_{\text{distributions of }\{\alpha\} \text{and}
\{\beta\}}\Sl_{\sigma_L\in
OP(\{\alpha_L\},\{\beta_L\}^T)}\Sl_{\sigma_R\in
OP(\{\alpha_R\},\{\beta_R\}^T)}(-1)^{n_{\beta}}A(H_1,...,H_m;\WH
g_1,\sigma_L|\sigma_R,\WH g_l).
 \eea
There are
$\Sl_{i=0}^{n_{\alpha}}\Sl_{j=0}^{n_{\beta}}\frac{(n_{\beta}-j+i)!}{i!(n_{\beta}-j)!}\frac{(n_{\alpha}-i+j)!}{j!(n_{\alpha}-i)!}
-\frac{(n_{\alpha}+n_{\beta})!}{n_{\alpha}!n_{\beta}!}$ terms in the
above equation. This is because the left set cannot only contain
$\WH g_l$. Respectively, there are only
$\frac{(n_{\alpha}+n_{\beta})!}{n_{\alpha}!n_{\beta}!}(n_{\alpha}+n_{\beta})$
terms in the BCFW expression $A(H_1,...,H_m;\WH g_1,[\sigma],\WH
g_l)$. Thus the numbers of terms also match. After summing all the
distributions of graviton, we get the KK relation (\ref{KK-mixed}).

%%%%%%%%%%%%%%%%%%%%%%%%%%%%%%%%%
\section{The disk relation for amplitudes with gluons coupled to one graviton}\label{Sect-disk}
%%%%%%%%%%%%%%%%%%%%%%%%%%%%%%%%%
In the sections \ref{Sect-C-O-R}, \ref{Sect-U(1)-D-I} and
\ref{Sect-KK}, we have discussed color-order reversed relation,
U(1)-decoupling identity and KK relation. These are the relations
among amplitudes with gluons coupled to gravitons at the leading
order. To study the relationship between gravity and gauge field, we
should also consider the relation between mixed amplitudes and pure
gluon amplitudes. In fact, the $(n,m) $ MHV amplitudes with only two
negative helicity gluons can be expressed by $n+2m$-point pure gluon
amplitudes \cite{Chen:2010sr}. This relation reflect the structure
of disk amplitudes in string theory \cite{Chen:2009tr},
\cite{Stieberger:2009hq}thus we call it disk relation. Though the
formula of the disk relation for MHV amplitudes given in
\cite{Chen:2010sr} depends on the helicity configuration, we expect
there is a helicity-independent formula like the KLT relation. In
\cite{Chen:2010sr}, a helicity independent formula of the disk
relation for $(n,1)$ amplitude is conjectured as
\bea\label{Ng1hrelation} A(g_1,g_2,...,g_n;H_1)=\Sl_{1<l\leq
n}s_{g_l,h_1}\Sl_{P}A_{n+2}(P), \eea
For a given $l$, $P$ are the permutations which preserve the
relative order of the gluons $g_1$, $g_2$,...,$g_n$. In $P$, one of
the gluon corresponding to $H_1$ is inserted between $g_1$, $g_l$
while the other one is inserted between $g_l$ and $g_1$. The
relation (\ref{Ng1hrelation}) can be rearranged into
\bea A(g_1,g_2,...,g_n;H_1)=\Sl_{1\leq i<j\leq
n}S_{ij}A(g_1,...,g_i,h_1,g_{i+1},...,g_j,\W
h_1,g_{j+1},...,g_n).~~~\Label{Disk-1} \eea
where $h_1$,  $\W h_1$ are the two gluons corresponding to the
graviton $H_1$ with momenta $p_{h_1}=p_{\W h_1}={1\over 2}p_{H_1}$,
$S_{ij} = \sum_{t=i+1}^{j} s_{g_t h_1}~$. We call this expression
{\bf the first formula} of the disk relation.

Let us give some remarks for formula (\ref{Disk-1}). First the
amplitude $A_{n+2}$ at the right hand side is not the general one
because $h_1,\W h_1$ always have same helicity and momentum, so the
right hand side is the sum of these constrained amplitudes and has
some properties which are not shared by unconstrained ones. Secondly
the pole structure is also constrained. For example,  we have
propagator $P= p_{g_1}+ p_{g_2}+...+p_{g_k}+p_{H_1}$ at the left
hand side, this is given by $(\W h_1,g_1,...,g_k,h_1)$ propagator at
the right hand side. Exactly because we need to match up all
possible propagators at the left hand side, we need to insert at all
possible positions as indicated at right hand side. However, with
this configuration we have the propagator $\W P= p_{g_1}+...+p_{g_k}
+ p_{h_1}$ at the right hand side which can not be found at the left
hand side, thus we need to insert $S_{ij}$ factors to cancel the
un-physical pole.

 Let us give one example to demonstrate the cancelation of
 unphysical pole. The first formula of disk relation (\ref{Disk-1}) gives
\bea A(g_1,g_2,g_3;H_1)& = &  s_{h_1 g_2} A(g_1,h_1, g_2, \W h_1,
g_3)+ (s_{h_1 g_2}+s_{h_1 g_3}) A(g_1,h_1, g_2, g_3,\W h_1) + s_{h_1
g_3} A(g_1, g_2, h_1, g_3, \W h_1) .\eea
To simplify above formula we do following
\bean  A(g_1,g_2,g_3;H_1) & = & s_{h_1 g_2} A(g_1,h_1, g_2, \W h_1,
g_3)+ (s_{h_1 g_2}+s_{h_1 g_3}) A(\W h_1, g_1,h_1, g_2, g_3) +
s_{h_1 g_3} A(\W h_1, g_1, g_2, h_1, g_3)\nn
& = & s_{h_1 g_2} A(g_1,h_1, g_2, \W h_1, g_3)+ s_{h_1 \W h_1} A(\W
h_1, h_1, g_1, g_2, g_3)=s_{h_1 g_2} A(g_1,h_1, g_2, \W h_1,
g_3),\eean
where we have used the BCJ relation for the last two terms at the
first line. The reason that term $s_{h_1 \W h_1} A(\W h_1, h_1, g_1,
g_2, g_3)$ is zero should explain. For $h_1,\W h_1$ nearby and
taking the collinear limit $p_{h_1}=p_{\W h_1}$, amplitude $A_5(\W
h_1, h_1, g_1, g_2, g_3)$ will reach $A_4(H_1, g_1, g_2, g_3)$
multiplying by a divergent collinear factor as ${1\over \Spaa{h_1|\W
h_1}}$ or ${1\over \Spbb{h_1|\W h_1}}$ depending on the helicity.
Thus after multiplying $s_{h_1\W h_1}= \Spaa{h_1|\W h_1}
\Spbb{h_1|\W h_1}$ we get zero.

Now we see the factor $s_{h_1 g_2}$ get rid the unphysical pole
$s_{h_1 g_2}$, $s_{\W h_1 g_2}$ as we have claimed. Furthermore,
using momentum conservation $s_{g_1 g_3}= s_{h_1 g_2} +s_{\W h_1
g_2}+s_{\W h_1 h_1}=2 s_{h_1 g_2}$, we arrive
\bea A(g_1,g_2,g_3;H_1) & = & 2s_{g_1 g_3} A(g_1, h_1, g_2, \W h_1,
g_3) \eea
which up to a scaling factor two is given already in
\cite{Chen:2010sr}.

{\bf Another formula:} As we have remarked, the right hand side of
(\ref{Disk-1}) is constrained in momentum space, thus there exists a
simplification. Let us focus on these special terms with  $j=n$,
i.e., we have $ A(g_1,...,g_i,h_1,...,g_n,\W h_1)$. The sum of these
terms is following
\bea  & & \sum_{1\leq i\leq n-1} S_{in} A
(g_1,...,g_i,h_1,...,g_n,\W h_1) \nn
& = & \sum_{1\leq i\leq n-1} (S_{in}+ s_{h_1 \W h_1}) A
(g_1,...,g_i,h_1,...,g_n,\W h_1)-\sum_{1\leq i\leq n-1} s_{h_1 \W
h_1} A (g_1,...,g_i,h_1,...,g_n,\W h_1)\nn
& = & \sum_{1\leq i\leq n-1} (S_{in}+ s_{h_1 \W h_1}) A
(g_1,...,g_i,h_1,...,g_n,\W h_1)+s_{h_1 \W h_1} [A
(g_1,...,g_n,h_1,\W h_1)+A(g_1,...,g_n,\W h_1, h_1)]\nn
& = & \sum_{1\leq i\leq n}  S_{i \W h_1} A
(g_1,...,g_i,h_1,...,g_n,\W h_1)+s_{h_1 \W h_1} A(g_1,...,g_n,\W
h_1, h_1),\eea
where in the third line we have used the $U(1)$-decoupling identity
for the second term in the second line, while in the fourth line we
have recombine them into the BCJ identity for the first term. For
the fourth line, the first term is zero by BCJ relation. The second
term is zero for which we have discussed already.
%\footnote{Our
%discussion true when and only when $n+2\geq 5$. The special case
%with $n+2=4$ is special because it reduces to three-point amplitude
%which is special.}.
% We need to use the
%property that when $h_1, \W h_1$ are collinear, the leading
%singularity is either ${1\over \Spaa{h_1|\W h_1}}$ or  ${1\over
%\Spbb{h_1|\W h_1}}$. Thus with the multiplicity of $s_{h_1 \W h_1}$
%we obtain zero.
This calculation leads us to {\bf the second formula}
\bea A (g_1,...,g_n; H_1) & = & \sum_{1\leq i<j\leq n-1} S_{ij}A(
g_1,..,g_i, h_1, g_{i+1},...,g_j, \W h_1,
g_{j+1},...,g_n).~~~\Label{Disk-2} \eea

{\bf Cyclic symmetry:} There is one point we have not discussed yet.
The two equations (\ref{Disk-1}) and (\ref{Disk-2}) have fixed the
$g_1$ at the first position, however, as the cyclic invariance of
the $A_{n;1}$, we should be able to use $g_i$ to write the expansion
at the right hand side. In other words, we should have
\bea A (g_1,...,g_n; H_1) & = & \sum_{2\leq i<j\leq n} S_{ij}A(
g_2,..,g_i, h_1, g_{i+1},...,g_j, \W h_1, g_{j+1},...,g_n, g_1)\nn
&+&\sum_{2\leq i\leq n} S_{i1}A( g_2,..,g_i, h_1,
g_{i+1},...,g_n,g_1,\W h_1).~~~\Label{Disk-Cyclic} \eea
To show (\ref{Disk-1}) is equal to (\ref{Disk-Cyclic}), we consider
the sum of different expressions. Using
\bean A (g_1,...,g_n; H_1) & = & \sum_{1\leq i<j\leq n} S_{ij}A(
g_1,..,g_i, h_1, g_{i+1},...,g_j, \W h_1, g_{j+1},...,g_n) \nn
& = & \sum_{2\leq i<j\leq n} S_{ij}A( g_1,g_2,..,g_i, h_1,
g_{i+1},...,g_j, \W h_1, g_{j+1},...,g_n)\nn & & + \sum_{2\leq j\leq
n} S_{1j}A( g_1,h_1,g_2,..,g_j, \W h_1, g_{j+1},...,g_n)\nn
& = & \sum_{2\leq i<j\leq n} S_{ij}A( g_2,..,g_i, h_1,
g_{i+1},...,g_j, \W h_1, g_{j+1},...,g_n,g_1)\nn & & + \sum_{2\leq
j\leq n} S_{1j}A( g_2,..,g_j, \W h_1, g_{j+1},...,g_n, g_1,
h_1)\eean
Using $s_{h_1 g_t}= s_{\W h_1 g_t}$, the sum of the second term in
previous equation can be simplified to
\bean s_{h_1 \W h_1}A( g_2,..,g_j, g_{j+1},...,g_n, g_1, \W h_1,
h_1)\eean
by BCJ relation. Thus, the definition (\ref{Disk-1}) gives
\bean A (g_1,...,g_n; H_1)& = & \sum_{2\leq i<j\leq n} S_{ij}A(
g_2,..,g_i, h_1, g_{i+1},...,g_j, \W h_1, g_{j+1},...,g_n,g_1)\nn &
& + s_{h_1 \W h_1}A_{n+2}( g_2,..,g_j, g_{j+1},...,g_n, g_1, \W h_1,
h_1).\eean
The definition (\ref{Disk-Cyclic}) gives
\bean A (g_1,...,g_n; H_1) & = & \sum_{2\leq i<j\leq n} S_{ij}A(
g_2,..,g_i, h_1, g_{i+1},...,g_j, \W h_1, g_{j+1},...,g_n,g_1)\nn &
& + \sum_{2\leq i\leq n} S_{i1}A( g_2,..,g_i, h_1,
g_{i+1},...,g_n,g_1,\W h_1)\nn & = & \sum_{2\leq i<j\leq n} S_{ij}A(
g_2,..,g_i, h_1, g_{i+1},...,g_j, \W h_1, g_{j+1},...,g_n,g_1)\nn &
& + s_{h_1\W h_1}A( g_2,..,g_n,g_1,h_1,\W h_1),\nn\eean
where again we have used the BCJ relation for the second term. Since
$h_1$, $\W h_1$ have same momentum and same helicity, they are
exchangeable, thus we have shown the cyclic property of our
definition\footnote{In fact, the second term with factor $s_{h_1 \W
h_1}$ is zero by our argument. However, our previous proof does not
need to use this observation.}.

{\bf One important remark:} Our proof starts with the contour
$I=\oint_{z=0} {dz \over z} \sum_{ij} S_{ij} A_{n+2}(z)=
A_{n+2}(z=0)$ around $z=0$ and we will show that contributions from
 finite poles are nothing, but  the BCFW
expansion of $A_{n;1}$. However, to show that $A_{n;1}=\sum S_{ij}
A_{n+2}$, we need to show the boundary contribution is zero.

For the boundary part, checking  our formula (\ref{Disk-1}) with
$(1,n)$ shift, we can see that when $\W h_1$ is not at the right
hand side of $g_n$, the factor $S_{ij}$ does not have
$z$-dependence, so we have ${1\over z}$ behavior. When $\W h_1$ is
at the right hand side of $g_n$, $S_{ij}\sim z$ but now the $g_1,
g_n$ are not nearby and have the ${1\over z^2}$ behavior. Overall we
have ${1\over z}$ behavior, so the boundary part is indeed zero.
This fact is also trivial in the second formula (\ref{Disk-2}) and
in fact, it has motivated us to do so.

%%%%%%%%%%%%%%%%%%%%%%%%%%%
\subsection{The proof of the second formula}
%%%%%%%%%%%%%%%%%%%%%%%%%%%
Now we consider the BCFW expansion of the right hand side of
(\ref{Disk-2}) with deformation of pair $(1,n)$. The good point of
formula (\ref{Disk-2}) is that the $g_1, g_n$ are always nearby, so
the BCFW expansion is much simpler.

Assuming the left hand part with gluons $(\WH g_1,...,g_k)$ and
right hand part with gluons $(g_{k+1},...,\WH g_n)$ where $k$ can
choose from $1$ to $n-1$, there are following different
distributions:
\begin{itemize}

\item (A) Both $h_1, \W h_1$ are in the left hand part, i.e., we
will have
\bea & & \sum_{1\leq i<j\leq k} {S_{ij}A_{L}(\WH g_1,..,g_i, h_1,
g_{i+1},...,g_j, \W h_1, g_{j+1},...,g_k, \WH P) A_{R}(-\WH P,
g_{k+1},...,\WH g_n)\over s_{g_{k+1}... g_n}}\nn
& = & A (\WH g_1,...,g_k, \WH P; H_1) A(- \WH P, g_{k+1},...,\WH
g_n)\over s_{g_{k+1}... g_n}.\eea
It is worth to mention that the left hand side include the special
case  $A_L(g_1,...,g_k, \W h_1, -P)$ and we have used (\ref{Disk-2})
for the left hand part.

\item (B) Both $h_1, \W h_1$ are in the right hand part, i.e., we
will have
\bea & & \sum_{k\leq i<j\leq n-1} {S_{ij}A_{L}(\WH g_1,..,g_k, \WH
P) A_{R}(-\WH P, g_{k+1},... g_i, h_1, ...,g_j, \W h_1, ...,\WH
g_n)\over s_{g_1...g_k}}\nn
& = &{ A (\WH g_1,...,g_k, \WH P)A(- \WH P, g_{k+1},...,\WH g_n;
H_1)\over s_{g_1...g_k}}.\eea
It is worth to mention that the right hand side include the special
case $A_R(-\WH P, h_1, g_{k+1},...,g_j, \W h_1, ...,\WH g_n)$.

\item (C) The last case is that $h_1$ belongs to the left hand part while
$\W h_1$ belongs to the right hand part. This case should give zero
contribution. We have
\bea I_C & = & \sum_{1\leq i\leq k} \sum_{k\leq j\leq n-1} S_{ij}
{A_L(\WH g_1,...,g_i, h_1,...,g_k,\WH P) A_R (-\WH P,
g_{k+1},...,g_j,...,\WH g_n)\over s_{12...k
h_1}}.~~~\Label{IC-1}\eea
In (\ref{IC-1}) when $i=k$ we have $A_L(...,g_k,h_1, \WH P)$ while
when $j=k$ we have $A_R(-\WH P, \W h_1, g_{k+1},...)$. Thus
rigorously we should exclude the case $i=j=k$. However, in this
special case, $S_{ij}=0$ so it does not, in fact, affect anything.

We observe that we can write
\bean S_{ij} & = & s_{h_1 g_{i+1}}+s_{h_1 g_{i+2}}+....+s_{h_1
g_j}\nn
& = & ( s_{h_1 g_{i+1}}+s_{h_1 g_{i+2}}+...+ s_{h_1 g_k}+s_{h_1 \WH
P})+ (s_{\W h_1 (-\WH P)}+ s_{\W h_1 g_{k+1}}+... s_{\W h_1 g_j})\nn
& \equiv  & S_{iL}+ S_{j R},\eean
where we have used the $p_{h_1}=p_{\W h_1}$ and $s_{h_1 \WH P}+s_{\W
h_1 (-\WH P)}=0$. Putting this back to (\ref{IC-1})  we obtain
\bea I_C & = &   \sum_{k\leq j\leq n-1} {[\sum_{1\leq i\leq k}
S_{iL}A_L(\WH g_1,...,g_i, h_1,...,g_k,\WH P)]  A_R (-\WH P,
g_{k+1},...,g_j,...,\WH g_n)\over s_{12...k h_1}}\nn
& + &  \sum_{1\leq i\leq k} {A_L(\WH g_1,...,g_i, h_1,...,g_k,\WH P)
[\sum_{k\leq j\leq n-1} S_{jR} A_R (-\WH P, g_{k+1},...,g_j,...,\WH
g_n)]\over s_{12...k h_1}}.~~~\Label{IC-2}\eea
As explained after the (\ref{IC-1}) for the special cases $i=j=k$,
the sum of these two lines in (\ref{IC-2}) will have coefficient
$S_{iL}+S_{jR}=s_{h_1 \WH P}+s_{\W h_1 (-\WH P)}=0$, thus  we can
include this special case.

With this writing, it is easy to see that each term is zero in
(\ref{IC-2}) by the BCJ relation. This zero result is nothing, but
our previous discussions about the cancelation of un-physical pole
with only $h_1$ or $\W h_1$.

Summing all the contributions, we finished the proof of  the second
formula of disk relation (\ref{Disk-2}).

\end{itemize}
%%%%%%%%%%%%%%%%%%%%%%%%%%%
\subsection{The first formula}
%%%%%%%%%%%%%%%%%%%%%%%%%%%

Since we have given the proof for the second formula, the first
formula must be true. However, we can also give similar proof for
the first formula, which demonstrate some new points. The trouble
part of the first formula (\ref{Disk-1}) is that when $j=n$, we will
have $A (g_1,..., g_n,h_1)$, i.e., the $g_1, g_n$ are not nearby
anymore, then the $\W h_1$ can be with $g_1$ as well as $g_n$.

Assuming the left hand part with gluons $(\WH g_1,...,g_k)$ and
right hand part with gluons $(g_{k+1},...,\WH g_n)$ where $k$ can
choose from $k=1$ to $k=n-1$, there are following different
distributions:
\begin{itemize}

\item (A) Both $h_1, \W h_1$ are in the left hand part. As we have mentioned, there are two
cases. The first one is
\bea I_{A-1} & =& \sum_{1\leq i<j\leq k} {S_{ij}A_{L}(\WH
g_1,..,g_i, h_1, g_{i+1},...,g_j, \W h_1, g_{j+1},...,g_k, \WH P)
A_{R}(-\WH P, g_{k+1},...,\WH g_n)\over s_{g_{k+1}... g_n}}\nn
& = &{ A (\WH g_1,...,g_k, \WH P; H_1) A(- \WH P, g_{k+1},...,\WH
g_n)\over s_{g_{k+1}... g_n}}.\eea

The second one is when $j=n$ and the BCFW expansion putting $\W h_1$
to the left hand part. Using the momentum conservation we can write
it as
\bea I_{A-2} & =&-\sum_{1\leq i\leq k} (\sum_{t=1}^{i-1} s_{g_t
h_1}){A_{L}(\W h_1, \WH g_1,..,g_i, h_1...,g_k, \WH P) A_{R}(-\WH P,
g_{k+1},...,\WH g_n)\over s_{g_{k+1}... g_n}},\eea
where if $i=1$ the factor is $s_{\WH g_1 h_1}$,i.e., we use the
shifted momentum of $g_1$. The sum over $i$ is not zero by BCJ
relation, but
\bea I_{A-2} & =& {-(s_{h_1 \W h_1})A_{L}( \WH g_1,..,g_k, \WH P,
h_1,\W  h_1) A_{R}(-\WH P, g_{k+1},...,\WH g_n)\over s_{g_{k+1}...
g_n}}.\eea
If we use the further condition $s_{h_1 \W h_1}=0$ we get
$I_{A-2}=0$.

\item (B) Both $h_1, \W h_1$ are in the right hand part, i.e., we
will have
\bea & & \sum_{k\leq i<j\leq n} {S_{ij}A_{L}(\WH g_1,..,g_k, \WH P)
A_{R}(-\WH P, g_{k+1},... g_i, h_1, ...,g_j, \W h_1, ...,\WH
g_n)\over s_{g_1...g_k}}\nn
& = &{ A (\WH g_1,...,g_k, \WH P) A(- \WH P, g_{k+1},...,\WH g_n;
H_1)\over s_{g_1...g_k}}.\eea

\item (C) The third case is that $h_1$ in the left hand part while
$\W h_1$ in the right hand part.  We have
\bea I_C & = & \sum_{1\leq i\leq k} \sum_{k\leq j\leq n} S_{ij}
{A_L(\WH g_1,...,g_i, h_1,...,g_k,\WH P) A_R (-\WH P,
g_{k+1},...,g_j,...,\WH g_n)\over s_{12...k
h_1}}.~~~\Label{2IC-1}\eea
In (\ref{2IC-1}) when $i=k$ we have $A_L(...,g_k,h_1, \WH P)$ while
when $j=k$ we have $A_R(-\WH P, \W h_1, g_{k+1},...)$. Again we
write $S_{ij} =   S_{iL}+ S_{j R}$ and put it back to (\ref{2IC-1})
to obtain
\bea I_C & = &   \sum_{k\leq j\leq n} {[\sum_{1\leq i\leq k}
S_{iL}A_L(\WH g_1,...,g_i, h_1,...,g_k,\WH P)]  A_R (-\WH P,
g_{k+1},...,g_j,...,\WH g_n)\over s_{12...k h_1}}\nn
& + &  \sum_{1\leq i\leq k} {A_L(\WH g_1,...,g_i, h_1,...,g_k,\WH P)
[\sum_{k\leq j\leq n} S_{jR} A_R (-\WH P, g_{k+1},...,g_j,...,\WH
g_n)]\over s_{12...k h_1}}.~~~\Label{2IC-2}\eea

The first term in (\ref{2IC-2}) will be zero by the BCJ relation.
For the second term, the sum $\sum_{k\leq j\leq n-1}$ is the BCJ
relation and it is zero. For the case $j=n$, the factor is given by
$s_{\W h_1 (-\WH P)}+s_{\W h_1 g_{k+1}}+...+s_{\W h_1 \W g_n}=-s_{\W
h_1 \W h_1}=0$ by momentum conservation. Thus the second term is
zero too.

\item (D) Comparing to previous subsection, we have addition case
where $\W h_1$ is in the left hand part while the $h_1$ is in the
right hand part. This can happen when and only when $j=n$, thus we
have following expression
\bea I_{D} & = & \sum_{k\leq i\leq n-1} S_{i\WH n} { A_L (\W h_1,
\WH g_1,...,g_k,\WH P) A_R (-\WH P, g_{k+1},...,g_i, h_1,...,\WH
g_n)\over P^2},\eea
where we have used the $\WH n$ to emphasize the shifted momentum.
One good thing of this part is that the sum over $i$ gives exact BCJ
relation, thus it is zero.

Thus, after considering all contributions, we have given the proof
for  the first formula of disk relation (\ref{Disk-1}).

\end{itemize}

%%%%%%%%%%%%%%%%%%%%%%%%%%%%%%%%
\section{From disk relation to color-order reversed relation}\label{Sect-C-O-R-disk}
%%%%%%%%%%%%%%%%%%%%%%%%%%%%%%%%
In the previous sections, we have seen two kinds of relations for
amplitudes with gluons coupled to gravitons:
\begin{itemize}
\item{(1)} Color-order reversed relation, $U(1)$-decoupling identity and KK relation which are the relations among the mixed amplitudes.
\end{itemize}
\begin{itemize}
\item{(2)} Disk relation which is the relations between mixed amplitudes and pure gluon amplitudes.
\end{itemize}
The disk relation express the $(n,m)$ amplitudes by $n+2m$-point
pure gluon amplitudes while the pure gluon amplitudes satisfy
color-order reversed relation, $U(1)$-decoupling identity and KK
relation. Thus we can give another derivation of the relations among
mixed amplitudes by using disk relation and the relations among pure
gluon amplitudes. In this section and the following two sections, we
will derive the relations among the amplitudes with gluons coupled
to one graviton in this way.
%%%%%%%%%%%%%%%%%%%%%%%%%%%
\subsection{An example}
%%%%%%%%%%%%%%%%%%%%%%%%%%%
We give the the color-order reversed relation for $(3,1)$ amplitudes
using disk relation as an example
\bea A(g_1,g_2,g_3;H_1)&=&s_{g_2h_1}A(g_1,h_1,g_2,\W h_1,g_3)\nn
&=&s_{g_2h_1}(-1)^5A(g_3,h_1,g_2,\W h_1,g_1)
=(-1)^3A(g_3,g_2,g_1;H_1). \eea
here the second formula of disk relation, the cyclic symmetry and
the invariance of the amplitudes under $h_1 \Leftrightarrow \W h_1$
have been used.
%%%%%%%%%%%%%%%%%%%%%%%%%%%
\subsection{General color-reversed relation for amplitudes with gluons coupled to one graviton}
%%%%%%%%%%%%%%%%%%%%%%%%%%%
From the second formula of disk relation, we can get the disk
relation \bean A(g_1,g_2,...,g_n;H_1) &=&\Sl_{2\leq i<j\leq
n-1}S_{ij}A(g_1,...,g_i,h_1,g_{i+1},...,g_j,\W
h_1,g_{j+1},...,g_n)\nn &=&\Sl_{2\leq i<j\leq
n-1}(-1)^{n+2}S_{ij}A(g_n,...,g_{j+1},\W h_1,g_{j},...,g_{i+1},
h_1,g_{i},...,g_1)\nn &=&(-1)^nA(g_n,...,g_1;H_1), \eean where we
have used the symmetry under $h_1\Leftrightarrow\W h_1$.

%%%%%%%%%%%%%%%%%%%%%%%%%%%%%%%%
\section{From disk relation to $U(1)$-decoupling identity}\label{Sect-U(1)-D-I-disk}
%%%%%%%%%%%%%%%%%%%%%%%%%%%%%%%%
In this section, we consider the $U(1)$ decoupling identity for
amplitudes with gluons coupled to one graviton by disk relation and
corresponding properties of pure gluon amplitudes.
%%%%%%%%%%%%%%%%%%%%%%%%%%%
\subsection{An example}
%%%%%%%%%%%%%%%%%%%%%%%%%%%
The simplest example is the $U(1)$-decoupling identity for $(3,1)$
amplitudes, this is same with the color-order reversed relation
given in the previous section. We will give one more example: the
$U(1)$-decoupling identity for $(4,1)$ amplitudes.

Using the first formula of disk relation, the amplitude
$A(g_1,g_2,g_3,g_4;H_1)$ can be given as sum of three parts \bean
&&A(g_1,g_2,g_3,g_4;H_1)=\mathbb{A}+\mathbb{B}+\mathbb{C} \eean
where \bea \mathbb{A}&=&s_{h_1g_2}A(g_1,h_1,g_2,\W h_1,g_3,g_4)\nn
\mathbb{B}&=&(s_{h_1g_2}+s_{h_1g_3})A(g_1,h_1,g_2,g_3,\W
h_1,g_4)+(s_{h_1g_2}+s_{h_1g_3}+s_{h_1g_4})A(g_1,h_1,g_2,g_3,g_4,\W
h_1)\nn &+&s_{h_1g_3}A(g_1,g_2,h_1,g_3,\W
h_1,g_4)+(s_{h_1g_3}+s_{h_1g_4})A(g_1,g_2,h_1,g_3,g_4,\W h_1)\nn
\mathbb{C}&=&s_{h_1g_4}A(g_1,g_2,g_3,h_1,g_4,\W h_1). \eea The
$\mathbb{A}$ part is the contribution with both $h_1$ and $\W h_1$
between $g_1$ and $g_3$. The $\mathbb{B}$ part is the contribution
with $h_1$ between $g_1$ and $g_3$, $\W h_1$ between $g_3$ and
$g_1$, while the $\mathbb{C}$ part is the contribution with both
$h_1$ and $\W h_1$ between $g_3$ and $g_1$.

With BCJ relation, $\mathbb{B}$ can be expressed by
$A(g_1,g_2,g_3,h_1,g_4,\W h_1)$ and $A(g_1,g_2,g_3,\W h_1,g_4,h_1)$
\bean &&(s_{h_1g_2}+s_{h_1g_3})A(g_1,h_1,g_2,g_3,\W
h_1,g_4)+s_{h_1g_3}A(g_1,g_2,h_1,g_3,\W h_1,g_4)\nn
&=&-(s_{h_1g_1}+s_{h_1g_2}+s_{h_1g_3})A(g_1,g_2,g_3,\W
h_1,g_4,h_1)-(s_{h_1g_1}+s_{h_1g_2}+s_{h_1g_3}+s_{h_1g_4})A(g_1,g_2,g_3,\W
h_1,h_1,g_4)\nn &=&s_{h_1g_4}A(g_1,g_2,g_3,\W h_1,g_4,h_1)+s_{h_1\W
h_1}A(g_1,g_2,g_3,\W h_1,h_1,g_4)\nn &=&s_{h_1g_4}A(g_1,g_2,g_3,\W
h_1,g_4,h_1). \eean \bean
&&(s_{h_1g_2}+s_{h_1g_3}+s_{h_1g_4})A(g_1,h_1,g_2,g_3,g_4,\W
h_1)+(s_{h_1g_3}+s_{h_1g_4})A(g_1,g_2,h_1,g_3,g_4,\W h_1)\nn
&=&-(s_{h_1g_1}+s_{h_1g_2}+s_{h_1g_3}+s_{h_1g_4})A(g_1,g_2,g_3,g_4,\W
h_1,h_1)-s_{h_1g_4}A(g_1,g_2,g_3,h_1,g_4,\W h_1)\nn &=&s_{h_1\W
h_1}A(g_1,g_2,g_3,g_4,\W h_1,h_1)-s_{h_1g_4}A(g_1,g_2,g_3,h_1,g_4,\W
h_1)\nn &=&-s_{h_1g_4}A(g_1,g_2,g_3,h_1,g_4,\W h_1). \eean

Thus \bean \mathbb{B}&=&s_{h_1g_4}A(g_1,g_2,g_3,\W
h_1,g_4,h_1)-s_{h_1g_4}A(g_1,g_2,g_3,h_1,g_4,\W h_1)=0. \eean

With $U(1)$-decoupling identity for pure gluon amplitudes,
$\mathbb{A}$  can be reexpressed as \bean &&\mathbb{A}\nn
&=&s_{h_1g_2}A(g_1,h_1,g_2,\W h_1,g_3,g_4)\nn
&=&s_{hg_2}[-A(g_1,g_4,h_1,g_2,\W h_1,g_3)-A(g_1,h_1,g_4,g_2,\W
h_1,g_3)-A(g_1,h_1,g_2,g_4,\W h_1,g_3)-A(g_1,h_1,g_2,\W
h_1,g_4,g_3)]. \eean

With KK relation for pure gluon amplitudes with $\{\alpha\}=\{g_2\}$
and $\{\beta\}=\{h_1,g_4,\W h_1\}$, $\mathbb{C}$ can be reexpressed
as \bean &&\mathbb{C}\nn &=&s_{h_1g_4}A(g_1,g_2,g_3,h_1,g_4,\W
h_1)\nn &=&s_{h_1g_4}[-A(g_1,g_2,\W h_1,g_4,h_1,g_3)-A(g_1,\W
h_1,g_2,g_4,h_1,g_3)-A(g_1,\W h_1,g_4,g_2,h_1,g_3)-A(g_1,\W
h_1,g_4,h_1,g_2,g_3)]\nn &=&s_{h_1g_4}[-A(g_1,g_2,h_1,g_4,\W
h_1,g_3)-A(g_1,h_1,g_2,g_4,\W h_1,g_3)-A(g_1,h_1,g_4,g_2,\W
h_1,g_3)-A(g_1,h_1,g_4,\W h_1,g_2,g_3)]. \eean Thus
$\mathbb{A}+\mathbb{C}$ becomes \bean &&\mathbb{A}+\mathbb{C}\nn
&=&-s_{h_1g_4}A(g_1,h_1,g_4,\W
h_1,g_2,g_3)-(s_{h_1g_2}+s_{hg_4})A(g_1,h_1,g_4,g_2,\W
h_1,g_3)-s_{h_1g_2}A(g_1,g_4,h_1,g_2,\W h_1,g_3)\nn
&-&s_{h_1g_2}A(g_1,h_1,g_2,\W
h_1,g_4,g_3)-(s_{h_1g_2}+s_{hg_4})A(g_1,h_1,g_2,g_4,\W
h_1,g_3)-s_{h_1g_4}A(g_1,g_2,h_1,g_4,\W h_1,g_3)\nn
&=&-A(g_1,g_4,g_2,g_3;H_1)-A(g_1,g_2,g_4,g_3;H_1). \eean At last we
have the $U(1)$-decoupling identity \bean
A(g_1,g_2,g_3,g_4;H_1)=-A(g_1,g_4,g_2,g_3;H_1)-A(g_1,g_2,g_4,g_3;H_1).
\eean

In the proof given above, the amplitude $A(g_1,g_2,g_3,g_4;H)$ is
expressed by six-gluon amplitudes via disk relation. These pure
gluon amplitudes can be given by $A(g_1,\{\alpha\},g_3,\{\beta\})$.
The sums of terms with $h,\W h_1\in\{\alpha\}$, $h,\W
h_1\in\{\beta\}$ and $h\in\{\alpha\},\W h_1\in\{\beta\}$ are
corresponding to $\mathbb{A}$, $\mathbb{C}$ and $\mathbb{B}$.
$\mathbb{B}$ term vanishes due to fundamental BCJ relation. With
$U(1)$-decoupling identity and KK-relation for pure gluon
amplitudes, the sum of $\mathbb{A}$ and $\mathbb{C}$ can give the R.
H. S. of the $U(1)$-decoupling identity for amplitudes with three
gluons coupled to one graviton.

%%%%%%%%%%%%%%%%%%%%%%%%%%%
\subsection{General $U(1)$-decoupling identity for amplitudes with gluons coupled to one graviton}
%%%%%%%%%%%%%%%%%%%%%%%%%%%
General $U(1)$-decoupling identity for gluons coupled to one
graviton can be given similarly. From disk relation we can see \bea
&&A(g_1,g_2,...,g_{n-1},g_n;H_1)\nn &=&\Sl_{1\leq
i<j<n-1}S_{ij}A(g_1,...,g_i,h_1,g_{i+1},...,g_j,\W
h_1,g_{j+1},...,g_{n-1},g_n)\nn &+&\Sl_{1\leq
i<n-1}S_{i,n-1}A(g_1,...,g_i,h_1,g_{i+1},...,,g_{n-1},\W h_1,g_n)\nn
&+&\Sl_{1\leq
i<n-1}S_{in}A(g_1,...,g_i,h_1,g_{i+1},...,g_{n-1},g_n,\W h_1)\nn
&+&S_{n-1,n}A(g_1,...,g_{n-1},h_1,g_n,\W h_1).\label{disk-u1} \eea
We can use $U(1)$-decoupling identity to express the first line of
the equation above as \bea &&\Sl_{1\leq
i<j<n-1}S_{ij}A(g_1,...,g_i,h_1,g_{i+1},...,g_j,\W
h_1,g_{j+1},...,g_{n-1},g_n)\nn &=&-\Sl_{1\leq
i<j<n-1}S_{ij}\Sl_{g_k\in\{g_1,...,g_i\}}A(g_1,...,g_k,g_n,g_{k+1},...,g_i,h_1,g_{i+1},...,g_j,\W
h_1,g_{j+1},...,g_{n-1})\nn &-&\Sl_{1\leq i\leq
j<n-1}S_{ij}\Sl_{g_k\in\{h_1,g_{i+1},...,g_j\}}A(g_1,...,g_i,h_1,g_{i+1},...,g_k,g_n,g_{k+1},...g_j,\W
h_1,g_{j+1},...,g_{n-1})\nn &-&\Sl_{1\leq
i<j<n-1}S_{ij}\Sl_{g_k\in\{\W
h_1,g_{j+1},...,g_{n-2}\}}A(g_1,...,g_i,h_1,g_{i+1},...,g_j,\W
h_1,g_{j+1},...,g_k,g_n,g_{k+1},...,g_{n-1}). \eea The three lines
correspond to contributions of different position of $g_N$: the
first line give the terms with $g_N$ on the left side of $h_1$, the
second lines give the terms with $g_N$ between $h_1$ and $\W h_1$
while the third line gives terms with $g_N$ on the right side of $\W
h_1$.

With KK relation for pure gluon amplitudes, the fourth line of
(\ref{disk-u1}) can be expressed as \bea
&&S_{n-1,n}A(g_1,...,g_{n-1},h_1,g_n,\W h_1)\nn &=&-\Sl_{1\leq i\leq
j<n-1}s_{h_1g_n}\Sl_{g_k\in\{h_1,g_{i+1},...,g_j\}}A(g_1,...,g_i,\W
h_1,g_{i+1},...,g_k,g_n,g_{k+1},...g_j,h_1,g_{j+1},...,g_{n-1}).
\eea Thus the sum of the first and the fourth lines is \bea
&&-\Sl_{1\leq
i<j<n-1}S_{ij}\Sl_{g_k\in\{g_1,...,g_i\}}A(g_1,...,g_k,g_n,g_{k+1},...,g_i,h_1,g_{i+1},...,g_j,\W
h_1,g_{j+1},...,g_{n-1})\nn &-&\Sl_{1\leq i\leq
j<n-1}(S_{ij}+s_{g_nh_1})\Sl_{g_k\in\{h_1,g_{i+1},...,g_j\}}A(g_1,...,g_i,h_1,g_{i+1},...,g_k,g_n,g_{k+1},...g_j,\W
h_1,g_{j+1},...,g_{n-1})\nn &-&\Sl_{1\leq
i<j<n-1}S_{ij}\Sl_{g_k\in\{\W
h_1,g_{j+1},...,g_{n-2}\}}A(g_1,...,g_i,h_1,g_{i+1},...,g_j,\W
h_1,g_{j+1},...,g_k,g_n,g_{k+1},...,g_{n-1})\nn &=&-\Sl_{1\leq
k<n-1}\Sl_{k\leq
i<j<n-1}S_{ij}A(g_1,...,g_k,g_n,g_{k+1},...,g_i,h_1,g_{i+1},...,g_j,\W
h_1,g_{j+1},...,g_{n-1})\nn &-&\Sl_{1\leq k<n-1}\Sl_{1\leq i\leq k
\leq
j<n-1}(S_{ij}+s_{g_nh_1})A(g_1,...,g_i,h_1,g_{i+1},...,g_k,g_n,g_{k+1},...g_j,\W
h_1,g_{j+1},...,g_{n-1})\nn &-&\Sl_{1\leq k<n-1}\Sl_{1\leq i<j\leq
k}S_{ij}A(g_1,...,g_i,h_1,g_{i+1},...,g_j,\W
h_1,g_{j+1},...,g_k,g_n,g_{k+1},...,g_{n-1}). \eea There are two
special terms in the equation above should be noticed: When $k=n-2$,
there is no amplitude with $g_n$ on the left side of $h$. when
$k=1$, there is no amplitude with $g_n$ on the right side of $\W
h_1$. Thus, rigorously, we should sum over $1\leq k<n-2$ for
amplitudes with $g_n$ on the left side of $h$ and $2\leq k<n-1$ for
amplitudes with $g_n$ on the right side of $\W h_1$. However, with
$S_{ij}=0$ for $i=j$, the terms with $k=n-2$ and $k=1$ can also be
contained in the two cases, respectively. Using the second formula
of disk relation, the equation above becomes \bea &&-\Sl_{1\leq
k<n-1}A(g_1,...,g_k,g_n,g_{k+1},...,g_{n-1}). \eea

With BCJ relation for pure gluon amplitudes, the second line of
(\eqref{disk-u1}) can be reexpressed as \bea
-(s_{h_1g_{n-1}}+s_{h_1g_{n-2}}+...+s_{h_1g_1})A(g_1,...,g_{n-1},\W
h_1,g_n,h_1)=s_{h_1g_n}A(g_1,...,g_{n-1},\W h_1,g_n,h_1), \eea while
the third line of (\eqref{disk-u1}) can be reexpressed as \bea
-s_{h_1g_n}A(g_1,...,g_{n-1},h,g_n,\W h_1). \eea Using the
invariance of the amplitude under $h_1\Leftrightarrow \W h_1$, the
second and the third lines cancel out with each other. Thus we get
the $U(1)$-decoupling identity \bea
&&A(g_1,g_2,...,g_{n-1},g_n;H_1)\nn &=&-\Sl_{1\leq
k<n-1}A(g_1,...,g_k,g_n,g_{k+1},...,g_{n-1})\nn
&=&-A(g_1,g_n,g_2,...,g_{n-1};H_1)-A(g_1,g_2,g_n,g_3,...,g_{n-1};H_1)-...-A(g_1,g_2,...,g_{n},g_{n-1};H_1).
\eea
%%%%%%%%%%%%%%%%%%%%%%%%%%%%%%%%
\section{From disk relation to KK relation}\label{Sect-KK-disk}
%%%%%%%%%%%%%%%%%%%%%%%%%%%%%%%%

As we have remarked, the color-order reversed relation and
$U(1)$-decoupling identity can be considered as two special cases of
KK relation. In this section, we will focus on the proof of KK
relation.
%%%%%%%%%%%%%%%%%%%%%%%%%%%
\subsection{An example}
%%%%%%%%%%%%%%%%%%%%%%%%%%%
We first prove the KK relation for $A(g_1,g_2,g_3,g_4,g_5,g_6;H_1)$
as an example. There are three different cases:
$A(g_1,\{g_2\},g_3,\{g_4,g_5,g_6\};H_1)$,
$A(g_1,\{g_2,g_3\},g_4,\{g_5,g_6\};H_1)$ and
$A(g_1,\{g_2,g_3,g_4\},g_5,\{g_6\};H_1)$. The first case is  a
$U(1)$-decoupling identity combined with a color-reversed relation.
The third case is just a $U(1)$-decoupling identity. Thus we only
consider $A(g_1,\{g_2,g_3\},g_4,\{g_5,g_6\};H_1)$ as the first
nontrivial case.

With the first formula of disk relation
\bea A(g_1,g_2,g_3,g_4,g_5,g_6;H_1)
&\equiv&\mathbb{A}+\mathbb{B}+\mathbb{C}, \eea
where
\bea
\mathbb{A}&=&s_{h_1g_2}A(g_1,h_1,g_2,\W
h_1,g_3,g_4,g_5,g_6)+(s_{h_1g_2}+s_{h_1g_3})A(g_1,h_1,g_2,g_3,\W
h_1,g_4,g_5,g_6)\nn &+&s_{h_1g_3}A(g_1,g_2,h_1,g_3,\W
h_1,g_4,g_5,g_6)\nn
\mathbb{C}&=&s_{h_1g_5}A(g_1,g_2,g_3,g_4,h_1,g_5,\W h_1,g_6)\nn
&+&(s_{h_1g_5}+s_{h_1g_6})A(g_1,g_2,g_3,g_4,h_1,g_5,g_6,\W h_1)\nn
&+&s_{h_1g_6}A(g_1,g_2,g_3,g_4,g_5,h_1,g_6,\W h_1)\nn
\mathbb{B}&=&(s_{h_1g_2}+s_{h_1g_3}+s_{h_1g_4})A(g_1,h_1,g_2,g_3,g_4,\W
h_1,g_5,g_6)\nn
&+&(s_{h_1g_2}+s_{h_1g_3}+s_{h_1g_4}+s_{h_1g_5})A(g_1,h_1,g_2,g_3,g_4,g_5,\W
h_1,g_6)\nn
&+&(s_{h_1g_2}+s_{h_1g_3}+s_{h_1g_4}+s_{h_1g_5}+s_{h_1g_6})A(g_1,h_1,g_2,g_3,g_4,g_5,g_6,\W
h_1)\nn &+&(s_{h_1g_3}+s_{h_1g_4})A(g_1,g_2,h_1,g_3,g_4,\W
h_1,g_5,g_6)\nn
&+&(s_{h_1g_3}+s_{h_1g_4}+s_{h_1g_5})A(g_1,g_2,h_1,g_3,g_4,g_5,\W
h_1,g_6)\nn
&+&(s_{h_1g_3}+s_{h_1g_4}+s_{h_1g_5}+s_{h_1g_6})A(g_1,g_2,h_1,g_3,g_4,g_5,g_6,\W
h_1)\nn &+&s_{h_1g_4}A(g_1,g_2,g_3,h_1,g_4,\W h_1,g_5,g_6)\nn
&+&(s_{h_1g_4}+s_{h_1g_5})A(g_1,g_2,g_3,h_1,g_4,g_5,\W h_1,g_6)\nn
&+&(s_{h_1g_4}+s_{h_1g_5}+s_{h_1g_6})A(g_1,g_2,g_3,h_1,g_4,g_5,g_6,\W
h_1). \eea
We divide all terms into these three categories according to the
positions of  $h_1$ and $\W h_1$: (1) category  $\mathbb{A}$
contains terms where  both $h_1$ and $\W h_1$ locate between $g_1$
and $g_4$; (2) category $\mathbb{B}$ contains terms where  $h_1$
locates between $g_1$ and $g_4$ while  $\W h_1$ locates  between
$g_4$ and $g_1$; (3) and finally category $\mathbb{C}$ contains term
where both $h_1$ and $\W h_1$ locate between $g_4$ and $g_1$.

With the BCJ relation for pure gluon amplitudes, the part
$\mathbb{B}$ is calculated as following:
\bea
\mathbb{B}&=&-(s_{h_1g_1}+s_{h_1g_2}+s_{h_1g_3}+s_{h_1g_4})A(g_1,g_2,g_3,g_4,\W
h_1,g_5,g_6,h_1)\nn
&-&(s_{h_1g_1}+s_{h_1g_2}+s_{h_1g_3}+s_{h_1g_4}+s_{h_1g_6})A(g_1,g_2,g_3,g_4,\W
h_1,g_5,h_1,g_6)\nn &-&s_{h_1g_5}A(g_1,g_2,g_3,g_4,h_1,g_5,\W
h_1,g_6)\nn
&-&(s_{h_1g_1}+s_{h_1g_2}+s_{h_1g_3}+s_{h_1g_4}+s_{h_1g_5})A(g_1,g_2,g_3,g_4,g_5,\W
h_1,g_6,h_1)\nn
&-&(s_{h_1g_5}+s_{h_1g_6})A(g_1,g_2,g_3,g_4,h_1,g_5,g_6,\W h_1)\nn
&-&s_{h_1g_6}A(g_1,g_2,g_3,g_4,g_5,h_1,g_6,\W h_1)\nn
&=&(s_{h_1g_5}+s_{h_1g_6})A(g_1,g_2,g_3,g_4,\W h_1,g_5,g_6,h_1)\nn
&+&s_{h_1g_5}A(g_1,g_2,g_3,g_4,\W h_1,g_5,h_1,g_6)\nn
&-&s_{h_1g_5}A(g_1,g_2,g_3,g_4,h_1,g_5,\W h_1,g_6)\nn
&+&s_{h_1g_6}A(g_1,g_2,g_3,g_4,g_5,\W h_1,g_6,h_1)\nn
&-&(s_{h_1g_5}+s_{h_1g_6})A(g_1,g_2,g_3,g_4,h_1,g_5,g_6,\W h_1)\nn
&-&s_{h_1g_6}A(g_1,g_2,g_3,g_4,g_5,h_1,g_6,\W h_1)\nn &=&0. \eea
where momentum conservation, $h_1\Leftrightarrow\W h_1$ and
collinear limits of pure gluon amplitudes have been used.

Each eight-gluon amplitude in $\mathbb{A}$ can be given by KK
relation for pure gluon amplitudes with four gluons in $\{\alpha\}$
and two gluons in $\{\beta\}$. Thus we have
 \bea
 \mathbb{A}&=&s_{h_1g_2}[A(g_1,g_6,g_5,h_1,g_2,\W h_1,g_3,g_4)
 +A(g_1,g_6,h_1,g_5,g_2,\W h_1,g_3,g_4)
 +A(g_1,g_6,h_1,g_2,g_5,\W h_1,g_3,g_4)\nn
 &+&A(g_1,g_6,h_1,g_2,\W h_1,g_5,g_3,g_4)
 +A(g_1,g_6,h_1,g_2,\W h_1,g_3,g_5,g_4)
 +A(g_1,h_1,g_6,g_5,g_2,\W h_1,g_3,g_4)\nn
 &+&A(g_1,h_1,g_6,g_2,g_5,\W h_1,g_3,g_4)
 +A(g_1,h_1,g_6,g_2,\W h_1,g_5,g_3,g_4)
 +A(g_1,h_1,g_6,g_2,\W h_1,g_3,g_5,g_4)\nn
 &+&A(g_1,h_1,g_2,g_6,g_5,\W h_1,g_3,g_4)
 +A(g_1,h_1,g_2,g_6,\W h_1,g_5,g_3,g_4)
 +A(g_1,h_1,g_2,g_6,\W h_1,g_3,g_5,g_4)\nn
 &+&A(g_1,h_1,g_2,\W h_1,g_6,g_5,g_3,g_4)
 +A(g_1,h_1,g_2,\W h_1,g_6,g_3,g_5,g_4)
 +A(g_1,h_1,g_2,\W h_1,g_3,g_6,g_5,g_4)]\nn
 &+&(s_{h_1g_2}+s_{h_1g_3})[A(g_1,g_6,g_5,h_1,g_2,g_3,\W h_1,g_4)
 +A(g_1,g_6,h_1,g_5,g_2,g_3,\W h_1,g_4)
 +A(g_1,g_6,h_1,g_2,g_5,g_3,\W h_1,g_4)\nn
 &+&A(g_1,g_6,h_1,g_2,g_3,g_5,\W h_1,g_4)
 +A(g_1,g_6,h_1,g_2,g_3,\W h_1,g_5,g_4)
 +A(g_1,h_1,g_6,g_5,g_2,g_3,\W h_1,g_4)\nn
 &+&A(g_1,h_1,g_6,g_2,g_5,g_3,\W h_1,g_4)
 +A(g_1,h_1,g_6,g_2,g_3,g_5,\W h_1,g_4)
 +A(g_1,h_1,g_6,g_2,g_3,\W h_1,g_5,g_4)\nn
 &+&A(g_1,h_1,g_2,g_6,g_5,g_3,\W h_1,g_4)
 +A(g_1,h_1,g_2,g_6,g_3,g_5,\W h_1,g_4)
 +A(g_1,h_1,g_2,g_6,g_3,\W h_1,g_5,g_4)\nn
 &+&A(g_1,h_1,g_2,g_3,g_6,g_5,\W h_1,g_4)
 +A(g_1,h_1,g_2,g_3,g_6,\W h_1,g_5,g_4)
 +A(g_1,h_1,g_2,g_3,\W h_1,g_6,g_5,g_4)]\nn
 &+&s_{h_1g_3}[A(g_1,g_6,g_5,g_2,h_1,g_3,\W h_1,g_4)
 +A(g_1,g_6,g_2,g_5,h_1,g_3,\W h_1,g_4)
 +A(g_1,g_6,g_2,h_1,g_5,g_3,\W h_1,g_4)\nn
 &+&A(g_1,g_6,g_2,h_1,g_3,g_5,\W h_1,g_4)
 +A(g_1,g_6,g_2,h_1,g_3,\W h_1,g_5,g_4)
 +A(g_1,g_2,g_6,g_5,h_1,g_3,\W h_1,g_4)\nn
 &+&A(g_1,g_2,g_6,h_1,g_5,g_3,\W h_1,g_4)
 +A(g_1,g_2,g_6,h_1,g_3,g_5,\W h_1,g_4)
 +A(g_1,g_2,g_6,h_1,g_3,\W h_1,g_5,g_4)\nn
 &+&A(g_1,g_2,h_1,g_6,g_5,g_3,\W h_1,g_4)
 +A(g_1,g_2,h_1,g_6,g_3,g_5,\W h_1,g_4)
 +A(g_1,g_2,h_1,g_6,g_3,\W h_1,g_5,g_4)\nn
 &+&A(g_1,g_2,h_1,g_3,g_6,g_5,\W h_1,g_4)
 +A(g_1,g_2,h_1,g_3,g_6,\W h_1,g_5,g_4)
 +A(g_1,g_2,h_1,g_3,\W h_1,g_6,g_5,g_4)].
 \eea

Each eight-gluon amplitude in $\mathbb{C}$ can be given by KK
relation for pure gluon amplitudes with two gluons in $\{\alpha\}$
and four gluons in $\{\beta\}$. Thus we have \bea
\mathbb{C}&=&s_{h_1g_5}[A(g_1,g_2,g_3,g_6,\W h_1,g_5,h_1,g_4)
+A(g_1,g_2,g_6,g_3,\W h_1,g_5,h_1,g_4) +A(g_1,g_2,g_6,\W
h_1,g_3,g_5,h_1,g_4)\nn &+&A(g_1,g_2,g_6,\W h_1,g_5,g_3,h_1,g_4)
+A(g_1,g_2,g_6,\W h_1,g_5,h_1,g_3,g_4) +A(g_1,g_6,g_2,g_3,\W
h_1,g_5,h_1,g_4)\nn &+&A(g_1,g_6,g_2,\W h_1,g_3,g_5,h_1,g_4)
+A(g_1,g_6,g_2,\W h_1,g_5,g_3,h_1,g_4) +A(g_1,g_6,g_2,\W
h_1,g_5,h_1,g_3,g_4)\nn &+&A(g_1,g_6,\W h_1,g_2,g_3,g_5,h_1,g_4)
+A(g_1,g_6,\W h_1,g_2,g_5,g_3,h_1,g_4) +A(g_1,g_6,\W
h_1,g_2,g_5,h_1,g_3,g_4)\nn &+&A(g_1,g_6,\W h_1,g_5,g_2,g_3,h_1,g_4)
+A(g_1,g_6,\W h_1,g_5,g_2,h_1,g_3,g_4) +A(g_1,g_6,\W
h_1,g_5,h_1,g_2,g_3,g_4)]\nn
&+&(s_{h_1g_5}+s_{h_1g_6})[A(g_1,g_2,g_3,\W h_1,g_6,g_5,h_1,g_4)
+A(g_1,g_2,\W h_1,g_3,g_6,g_5,h_1,g_4) +A(g_1,g_2,\W
h_1,g_6,g_3,g_5,h_1,g_4)\nn &+&A(g_1,g_2,\W h_1,g_6,g_5,g_3,h_1,g_4)
+A(g_1,g_2,\W h_1,g_6,g_5,h_1,g_3,g_4) +A(g_1,\W
h_1,g_2,g_3,g_6,g_5,h_1,g_4)\nn &+&A(g_1,\W
h_1,g_2,g_6,g_3,g_5,h_1,g_4) +A(g_1,\W h_1,g_2,g_6,g_5,g_3,h_1,g_4)
+A(g_1,\W h_1,g_2,g_6,g_5,h_1,g_3,g_4)\nn &+&A(g_1,\W
h_1,g_6,g_2,g_3,g_5,h_1,g_4) +A(g_1,\W h_1,g_6,g_2,g_5,g_3,h_1,g_4)
+A(g_1,\W h_1,g_6,g_2,g_5,h_1,g_3,g_4)\nn &+&A(g_1,\W
h_1,g_6,g_5,g_2,g_3,h_1,g_4) +A(g_1,\W h_1,g_6,g_5,g_2,h_1,g_3,g_4)
+A(g_1,\W h_1,g_6,g_5,h_1,g_2,g_3,g_4)]\nn
&+&s_{h_1g_6}[A(g_1,g_2,g_3,\W h_1,g_6,h_1,g_5,g_4) +A(g_1,g_2,\W
h_1,g_3,g_6,h_1,g_5,g_4) +A(g_1,g_2,\W h_1,g_6,g_3,h_1,g_5,g_4)\nn
&+&A(g_1,g_2,\W h_1,g_6,h_1,g_3,g_5,g_4) +A(g_1,g_2,\W
h_1,g_6,h_1,g_5,g_3,g_4) +A(g_1,\W h_1,g_2,g_3,g_6,h_1,g_5,g_4)\nn
&+&A(g_1,\W h_1,g_2,g_6,g_3,h_1,g_5,g_4) +A(g_1,\W
h_1,g_2,g_6,h_1,g_3,g_5,g_4) +A(g_1,\W
h_1,g_2,g_6,h_1,g_5,g_3,g_4)\nn &+&A(g_1,\W
h_1,g_6,g_2,g_3,h_1,g_5,g_4) +A(g_1,\W h_1,g_6,g_2,h_1,g_3,g_5,g_4)
+A(g_1,\W h_1,g_6,g_2,h_1,g_5,g_3,g_4)\nn &+&A(g_1,\W
h_1,g_6,h_1,g_2,g_3,g_5,g_4) +A(g_1,\W h_1,g_6,h_1,g_2,g_5,g_3,g_4)
+A(g_1,\W h_1,g_6,h_1,g_5,g_2,g_3,g_4)].
 \eea
Terms in $\mathbb{A}+\mathbb{C}$ can be rearranged by the different
permutations of gluons $g_1$,...,$g_6$. For example, for permutation
$g_1$, $g_6$, $g_5$, $g_2$, $g_3$, $g_4$, we have
\bea&&s_{h_1g_6}A(g_1,h_1,g_6,\W h_1,g_5,g_2,g_3,g_4)\nn
&+&(s_{h_1g_5}+s_{h_1g_6})A(g_1,h_1,g_6,g_5,\W h_1,g_2,g_3,g_4)\nn
&+&(s_{h_1g_6}+s_{h_1g_5}+s_{h_1g_2})A(g_1,h_1,g_6,g_5,g_2,\W
h_1,g_3,g_4)\nn
&+&(s_{h_1g_6}+s_{h_1g_5}+s_{h_1g_2}+s_{h_1g_3})A(g_1,h_1,g_6,g_5,g_2,g_3,\W
h_1,g_4)\nn &+&(s_{h_1g_5})A(g_1,g_6,h_1,g_5,\W h_1,g_2,g_3,g_4)\nn
&+&(s_{h_1g_5}+s_{h_1g_2})A(g_1,g_6,h_1,g_5,g_2,\W h_1,g_3,g_4)\nn
&+&(s_{h_1g_5}+s_{h_1g_2}+s_{h_1g_3})A(g_1,g_6,h_1,g_5,g_2,g_3,\W
h_1,g_4)\nn &+&(s_{h_1g_2})A(g_1,g_6,g_5,h_1,g_2,\W h_1,g_3,g_4)\nn
&+&(s_{h_1g_2}+s_{h_1g_3})A(g_1,g_6,g_5,h_1,g_2,g_3,\W h_1,g_4)\nn
&+&(s_{h_1g_3})A(g_1,g_6,g_5,g_2,h_1,g_3,\W h_1,g_4)\nn
&=&A(g_1,g_6,g_5,g_2,g_3,g_4;H_1).
 \eea
In the same way, other terms give $A(g_1,g_6,g_2,g_5,g_3,g_4;H_1)$,
$A(g_1,g_6,g_2,g_3,g_5,g_4;H_1)$, $A(g_1,g_2,g_6,g_5,g_3,g_4;H_1)$,
$A(g_1,g_2,g_6,g_3,g_5,g_4;H_1)$ and
$A(g_1,g_2,g_3,g_6,g_5,g_4;H_1)$. Thus we have
 \bea
\mathbb{A}+\mathbb{C}&=&A(g_1,g_2,g_3,g_4,g_5,g_6;H_1)\nn
&=&A(g_1,g_6,g_5,g_2,g_3,g_4;H_1)+A(g_1,g_6,g_2,g_5,g_3,g_4;H_1)
+A(g_1,g_6,g_2,g_3,g_5,g_4;H_1)\nn &+&A(g_1,g_2,g_6,g_5,g_3,g_4;H_1)
+A(g_1,g_2,g_6,g_3,g_5,g_4;H_1) +A(g_1,g_2,g_3,g_6,g_5,g_4;H_1).
 \eea
This is just the KK relation with two gluons $g_2$, $g_3$ in
$\{\alpha\}$ and $g_5$, $g_6$ in $\{\beta\}$.

This example has demonstrate the idea of general proof. We divide
all terms given by disk relation into three categories:
 $\mathbb{A}$, $\mathbb{B}$ and $\mathbb{C}$. $\mathbb{A}$,
contains the amplitudes with $h_1$, $\W h_1$ in set $\{\alpha\}$.
$\mathbb{B}$ contains the amplitudes with $h_1$, $\W h_1$ in sets
$\{\alpha\}$,$\{\beta\}$, respectively . $\mathbb{C}$ contains the
amplitudes with $h_1$, $\W h_1$ in set $\{\beta\}$. $\mathbb{B}$
vanishes due to BCJ relation. The sum of $\mathbb{A}$ and
$\mathbb{C}$ is just the R.H.S. of KK relation for amplitudes gluons
coupled to one graviton.

%%%%%%%%%%%%%%%%%%%%%%%%%%%
\subsection{General KK relation for amplitudes with gluons coupled to one graviton}
%%%%%%%%%%%%%%%%%%%%%%%%%%%

Having above example and the idea how to prove, now we turn to the
general discussion on KK relation. We consider the
$A(g_1,\{g_2,...,g_{l-1}\},g_l,\{g_{l+1},...,g_{n}\};H_1)$ case.
Disk relation express the amplitude with $n$ gluons coupled to one
graviton as
\bea &&A(g_1,g_2,...,g_n;H_1)\nn &=&\Sl_{1\leq
i<j<l}S_{ij}A(g_1,...,g_i,h_1,g_{i+1},...,g_j,\W
h_1,g_{j+1}...,g_l,...,g_n)\nn &+&\Sl_{l\leq i<j\leq
n}S_{ij}A(g_1,...,g_l,...,g_i,h_1,g_{i+1},...,g_j,\W
h_1,g_{j+1},...,g_n)\nn &+&\Sl_{1\leq i\leq l-1,l\leq k\leq
n}S_{ij}A(g_1,...,g_i,h_1,g_{i+1},...,g_l,...g_j,\W
h_1,g_{j+1},...,g_n)\label{disk-kk}. \eea
where three lines are categories $\mathbb{A}$, $\mathbb{C}$ and
$\mathbb{B}$ respectively.

BCJ relation for pure gluon amplitudes can always reexpress
$A(g_1,...,g_i,h_1,g_{i+1},...,g_l,...g_j,\W h_1,g_{j+1},...,g_n)$
in the third line by amplitudes \bea
A(g_1,...,g_l,...g_i,h_1,g_{i+1},...,g_j,\W h_1,g_{j+1},...,g_n)
\eea with a factor $-S_{ij}$ and \bea A(g_1,...,g_l,...,g_i,\W
h_1,g_{i+1},...,...g_j,h_1,g_{j+1},...,g_n) \eea with a factor
$S_{ij}$. Because of the invariance under $h_1\Leftrightarrow\W
h_1$, the third term of (\eqref{disk-kk}) vanishes. The cancellation
in the discussions on $U(1)$-decoupling identity is just a special
case.

The first line of (\eqref{disk-kk}) can be reexpressed by KK
relation for pure gluon amplitudes \bea \Sl_{1\leq
i<j<l}S_{ij}(-1)^{n-l}\Sl_{\sigma\in
OP(\{g_2,...,g_i,h_1,g_{i+1}...,g_j,\W
h_1,g_{j+1}...,g_{l-1}\},\{g_n,...,g_{l+1}\})}A(g_1,\sigma,g_l).
\eea This can be expressed in another way: Insert $g_{l+1},...,g_n$
at the positions between $g_1$ and $g_l$. The insertions should
reversed the relative order of $g_{l+1},...,g_n$, i.e.,
$g_n,...,g_1$. For a given insertion of $g_n,...,g_{l+1}$, we insert
$h_1$ at the positions between $g_1$ and $g_l$, then insert $\W h_1$
at the positions between $h_1$ and $g_l$. The factor $S_{ij}$ is
$S_{ij}=s_{h_1g_{i+1}}+,...,+s_{h_1g_{j}}$. $g_{i+1},...,g_{j}$ are
the gluons between $h_1$ and $\W h_1$. They come from the set
$\{g_2,...,g_{l-1}\}$. Multiplying the factor $(-1)^{n-l}$ and
summing over all the insertions of $g_{l+1},...,g_{n}$ with the
relative order $g_n,...,g_{l+1}$, the first line of
(\eqref{disk-kk}) becomes \bea\label{disk-kk-1}
(-1)^{n-l}\Sl_{\sigma\in
OP(\{g_2,...,g_{l-1}\},\{g_n,...,g_{l+1}\})}\Sl_{1\leq
i<j<l}S_{ij}A(g_1,...,h_1,...,\W h_1,...,g_n). \eea

The second line of (\eqref{disk-kk}) can be reexpressed by KK
relation for pure gluon amplitudes \bea \Sl_{l\leq i<j\leq
n}S_{ij}(-1)^{n-l+2}\Sl_{\sigma\in
OP(\{g_2,...,g_{l-1}\},\{g_n,...,g_{j+1},\W
h_1,g_j,...,g_{i+1},h_1,g_{i},...,g_{l+1}\})}A(g_1,\sigma,g_l). \eea
This can be expressed in a similar way as the first line: We insert
$g_{l+1},...,g_{n}$ at the positions between $g_1$ and $g_l$ with
the relative order $g_n,...,g_{l+1}$. For a given insertion of
$g_n,...,g_{l+1}$, we insert $\W h_1$ at the positions between $g_1$
and $g_l$, then insert $h_1$ at the positions between $\W h_1$ and
$g_l$. The factor $S_{ij}$ is
$S_{ij}=s_{h_1g_{i+1}}+,...,+s_{h_1g_{j}}$  where
$g_{i+1},...,g_{j}$ are the gluons between $h_1$ and $\W h_1$. They
come from the set $\{g_n,...,g_{l+1}\}$. Multiplying the factor
$(-1)^{n-l+2}=(-1)^{n-l}$ and summing over all the possible
insertions of $g_n,...,g_{l+1}$, the second line of
(\eqref{disk-kk}) becomes \bea\label{disk-kk-2}
(-1)^{n-l}\Sl_{\sigma\in
OP(\{g_2,...,g_{l-1}\},\{g_n,...,g_{l+1}\})}\Sl_{l\leq
i<j<n}S_{ij}A(g_1,...,h_1,...,\W h_1,...,g_n), \eea where the
invariance of the amplitudes under $h_1\Leftrightarrow\W h_1$ has
been used. For a given permutation $g_1,...,h_1,...,\W h_1,...,g_n$,
the sum of the factors $S_{ij}$ in (\eqref{disk-kk-1}) and
(\eqref{disk-kk-2}) becomes $s_{h_1g_{i+1}}+...,+s_{h_1g_j}$. Here
$g_{i+1}$,...,$g_j$ are all the gluons between $h_1$ and $\W h_1$.
Thus the sum of (\eqref{disk-kk-1}) and (\eqref{disk-kk-2}) just
gives KK relation.

%%%%%%%%%%%%%%%%%%%%%%%%%%%%%%%%%%
\section{Remarks}\label{Sect-remarks}
%%%%%%%%%%%%%%%%%%%%%%%%%%%%%%%%%%
In this paper, we have shown two kinds of amplitudes relations: the
relations among mixed amplitudes and the disk relation between mixed
amplitudes and pure gluon amplitudes. In the first case, the
color-order reversed relation, $U(1)$-decoupling identity and KK
relation for amplitudes with gluons coupled to gravitons at the
leading order have the similar  forms like those in pure gluon case,
i.e. the appearing of gravitons does not cause  any differences.

Unlike above three relations, the BCJ relation for mixed amplitudes
is more tricky. The basic reason is the  nontrivial kinematic
factors $s_{ij}$ multiplying these amplitudes. The momentum
conservation condition, which is crucial for the BCJ relation, will
be modified by gravitons, thus the naive BCJ relation will not hold.
There are two possibilities we can consider. The first one is to
modify these kinematic factors to make BCJ relation. The second
possibility is the the naive BCJ relation is not zero, but something
which can be written down in general. These two possibilities are
under investigation.

Another question is the explicit disk relations with gluons coupled
to more than one graviton. Although the string theory analysis tells
us this formula must exist, its explicit form has not been given
except the MHV amplitude\cite{Chen:2010sr}.

%%%%%%%%%%%%%%%%%%%%%%
\subsection*{Acknowledgements}
%%%%%%%%%%%%%%%%%%%%%%
Y. J. Du would like to thank Q. Ma, J. L. Li and Z. B. Zhang for
many helpful discussions. He would also like to thank KITPC for
hospitality. Y. X. Chen and Y. J. Du are supported in part by the
NSF of China Grant No. 10775116, No. 11075138, and 973-Program Grant
No. 2005CB724508. BF is supported by fund from Qiu-Shi, the
Fundamental Research Funds for the Central Universities with
contract number 2009QNA3015, as well as Chinese NSF funding under
contract No.10875104.

%%%%%%%%%%%%%%%%%%%%%%%%%%%%%%%%%%%%%%%%%%%%%%%%%%%%


\begin{thebibliography}{999}
%%%%%%%%%%%%%%%%%%%%

%%KK-BCJ relation

%\cite{Kleiss:1988ne}
\bibitem{Kleiss:1988ne}
  R.~Kleiss and H.~Kuijf,
  ``MULTI - GLUON CROSS-SECTIONS AND FIVE JET PRODUCTION AT HADRON COLLIDERS,''
  Nucl.\ Phys.\  B {\bf 312} (1989) 616.
  %%CITATION = NUPHA,B312,616;%%

%\cite{Bern:2008qj}
\bibitem{Bern:2008qj}
  Z.~Bern, J.~J.~M.~Carrasco and H.~Johansson,
  ``New Relations for Gauge-Theory Amplitudes,''
  Phys.\ Rev.\  D {\bf 78} (2008) 085011
  [arXiv:0805.3993 [hep-ph]].
  %%CITATION = PHRVA,D78,085011;%%

%%monodromy

%\cite{BjerrumBohr:2009rd}
\bibitem{BjerrumBohr:2009rd}
  N.~E.~J.~Bjerrum-Bohr, P.~H.~Damgaard and P.~Vanhove,
  ``Minimal Basis for Gauge Theory Amplitudes,''
  Phys.\ Rev.\ Lett.\  {\bf 103} (2009) 161602
  [arXiv:0907.1425 [hep-th]].
  %%CITATION = PRLTA,103,161602;%%

%\cite{Stieberger:2009hq}
\bibitem{Stieberger:2009hq}
  S.~Stieberger,
  ``Open \& Closed vs. Pure Open String Disk Amplitudes,''
  arXiv:0907.2211 [hep-th].
  %%CITATION = ARXIV:0907.2211;%%

%%field theory proof

 %\cite{DelDuca:1999rs}
\bibitem{DelDuca:1999rs}
  V.~Del Duca, L.~J.~Dixon and F.~Maltoni,
  ``New color decompositions for gauge amplitudes at tree and loop level,''
  Nucl.\ Phys.\  B {\bf 571} (2000) 51
  [arXiv:hep-ph/9910563].
  %%CITATION = NUPHA,B571,51;%%

%\cite{Feng:2010my}
\bibitem{Feng:2010my}
  B.~Feng, R.~Huang and Y.~Jia,
  ``Gauge Amplitude Identities by On-shell Recursion Relation in S-matrix
  Program,''
  arXiv:1004.3417 [hep-th].
  %%CITATION = ARXIV:1004.3417;%%

%\cite{Tye:2010kg}
\bibitem{Tye:2010kg}
  H.~Tye and Y.~Zhang,
  ``Comment on the Identities of the Gluon Tree Amplitudes,''
  arXiv:1007.0597 [hep-th].
  %%CITATION = ARXIV:1007.0597;%%

%\cite{Jia:2010nz}
\bibitem{Jia:2010nz}
  Y.~Jia, R.~Huang and C.~Y.~Liu,
  ``U(1)-decoupling, KK and BCJ relations in $\mathcal{N}=4$ SYM,''
  Phys.\ Rev.\  D {\bf 82} (2010) 065001
  [arXiv:1005.1821 [hep-th]].
  %%CITATION = PHRVA,D82,065001;%%


%%%%%%%%%%%%%%%%%%%%%%%%%%%%%%%%%%%%%%%%%%%%%%%

%%KLT relation


%\cite{Kawai:1985xq}
\bibitem{Kawai:1985xq}
  H.~Kawai, D.~C.~Lewellen and S.~H.~H.~Tye,
  ``A Relation Between Tree Amplitudes Of Closed And Open Strings,''
  Nucl.\ Phys.\  B {\bf 269} (1986) 1.
  %%CITATION = NUPHA,B269,1;%%

%%KLT field theory limit

%\cite{Berends:1988zp}
\bibitem{Berends:1988zp}
  F.~A.~Berends, W.~T.~Giele and H.~Kuijf,
  ``On relations between multi - gluon and multigraviton scattering,''
  Phys.\ Lett.\  B {\bf 211} (1988) 91.
  %%CITATION = PHLTA,B211,91;%%

  %\cite{Bern:1998ug}
\bibitem{Bern:1998ug}
  Z.~Bern, L.~J.~Dixon, D.~C.~Dunbar, M.~Perelstein and J.~S.~Rozowsky,
  ``On the relationship between Yang-Mills theory and gravity and its
  implication for ultraviolet divergences,''
  Nucl.\ Phys.\  B {\bf 530} (1998) 401
  [arXiv:hep-th/9802162].
  %%CITATION = NUPHA,B530,401;%%

%%%KLT proof

%\cite{BjerrumBohr:2010ta}
\bibitem{BjerrumBohr:2010ta}
  N.~E.~J.~Bjerrum-Bohr, P.~H.~Damgaard, B.~Feng and T.~Sondergaard,
  ``Gravity and Yang-Mills Amplitude Relations,''
  arXiv:1005.4367 [hep-th].
  %%CITATION = ARXIV:1005.4367;%%

  %\cite{BjerrumBohr:2010zb}
\bibitem{BjerrumBohr:2010zb}
  N.~E.~J.~Bjerrum-Bohr, P.~H.~Damgaard, B.~Feng and T.~Sondergaard,
  ``New Identities among Gauge Theory Amplitudes,''
  Phys.\ Lett.\  B {\bf 691} (2010) 268
  [arXiv:1006.3214 [hep-th]].
  %%CITATION = PHLTA,B691,268;%%

%\cite{BjerrumBohr:2010yc}
\bibitem{BjerrumBohr:2010yc}
  N.~E.~J.~Bjerrum-Bohr, P.~H.~Damgaard, B.~Feng and T.~Sondergaard,
  ``Proof of Gravity and Yang-Mills Amplitude Relations,''
  JHEP {\bf 1009} (2010) 067
  [arXiv:1007.3111 [hep-th]].
  %%CITATION = JHEPA,1009,067;%%


%\cite{Feng:2010br}
\bibitem{Feng:2010br}
  B.~Feng and S.~He,
  ``KLT and New Relations for N=8 SUGRA and N=4 SYM,''
  JHEP {\bf 1009} (2010) 043
  [arXiv:1007.0055 [hep-th]].
  %%CITATION = JHEPA,1009,043;%%


%KLT gravity-matter coupling

%\cite{Bern:1999bx}
\bibitem{Bern:1999bx}
  Z.~Bern, A.~De Freitas and H.~L.~Wong,
  ``On the coupling of gravitons to matter,''
  Phys.\ Rev.\ Lett.\  {\bf 84} (2000) 3531
  [arXiv:hep-th/9912033].
  %%CITATION = PRLTA,84,3531;%%

%\cite{BjerrumBohr:2004wh}
\bibitem{BjerrumBohr:2004wh}
  N.~E.~J.~Bjerrum-Bohr and K.~Risager,
  ``String theory and the KLT-relations between gravity and gauge theory
  including external matter,''
  Phys.\ Rev.\  D {\bf 70} (2004) 086011
  [arXiv:hep-th/0407085].
  %%CITATION = PHRVA,D70,086011;%%




%%KLT-loop


%\cite{KLT-loop}
\bibitem{KLT-loop}
 Z.~Bern, J.~J.~M.~Carrasco and H.~Johansson,
  ``Perturbative Quantum Gravity as a Double Copy of Gauge Theory,''
  Phys.\ Rev.\ Lett.\  {\bf 105} (2010) 061602
  [arXiv:1004.0476 [hep-th]];
   Z.~Bern, T.~Dennen, Y.~t.~Huang and M.~Kiermaier,
  ``Gravity as the Square of Gauge Theory,''
Phys.\ Rev.\  D {\bf 82} (2010) 065003
  [arXiv:1004.0693 [hep-th]];





%\cite{KLT-understanding}
\bibitem{KLT-understanding}
S.~H.~Henry Tye and Y.~Zhang,
  ``Dual Identities inside the Gluon and the Graviton Scattering Amplitudes,''
  JHEP {\bf 1006} (2010) 071
  [arXiv:1003.1732 [hep-th]];
  N.~E.~J.~Bjerrum-Bohr and P.~Vanhove,
  ``Monodromy and Kawai-Lewellen-Tye Relations for Gravity Amplitudes,''
  arXiv:1003.2396 [hep-th];
  N.~E.~J.~Bjerrum-Bohr, P.~H.~Damgaard, T.~Sondergaard and P.~Vanhove,
  ``Monodromy and Jacobi-like Relations for Color-Ordered Amplitudes,''
  JHEP {\bf 1006} (2010) 003
  [arXiv:1003.2403 [hep-th]];
  D.~Vaman and Y.~P.~Yao,
  ``Constraints and Generalized Gauge Transformations on Tree-Level Gluon and
  Graviton Amplitudes,''
  arXiv:1007.3475 [hep-th];
   H.~Elvang and M.~Kiermaier,
  ``Stringy KLT relations, global symmetries, and $E_7(7)$ violation,''
  arXiv:1007.4813 [hep-th];
  B.~Feng, S.~He, R.~Huang and Y.~Jia,
  ``Note on New KLT relations,''
  arXiv:1008.1626 [hep-th];
  Y.~Abe,
  ``Holonomies of gauge fields in twistor space 3: gravity as a square of N=4
  theory,''
  arXiv:1008.2800 [hep-th].
 N.~E.~J.~Bjerrum-Bohr, P.~H.~Damgaard, T.~Sondergaard and P.~Vanhove,
  ``The Momentum Kernel of Gauge and Gravity Theories,''
  arXiv:1010.3933 [hep-th].















%%BCFW
%\cite{Britto:2004ap}
\bibitem{Britto:2004ap}
  R.~Britto, F.~Cachazo and B.~Feng,
  ``New Recursion Relations for Tree Amplitudes of Gluons,''
  Nucl.\ Phys.\  B {\bf 715} (2005) 499
  [arXiv:hep-th/0412308].
  %%CITATION = NUPHA,B715,499;%%
%\cite{Britto:2005fq}
\bibitem{Britto:2005fq}
  R.~Britto, F.~Cachazo, B.~Feng and E.~Witten,
  ``Direct Proof Of Tree-Level Recursion Relation In Yang-Mills Theory,''
  Phys.\ Rev.\ Lett.\  {\bf 94} (2005) 181602              (2005) 181602
  [arXiv:hep-th/0501052].
  %%CITATION = PRLTA,94,181602;%%


%\cite{z-infinite}
\bibitem{z-infinite}
J.~Bedford, A.~Brandhuber, B.~J.~Spence and G.~Travaglini, ``A
recursion relation for gravity amplitudes,'' Nucl.\ Phys.\  B {\bf
721} (2005) 98 [arXiv:hep-th/0502146]; F.~Cachazo and P.~Svrcek,
``Tree level recursion relations in general relativity,''
arXiv:hep-th/0502160; N.~E.~J.~Bjerrum-Bohr, D.~C.~Dunbar, H.~Ita,
W.~B.~Perkins and K.~Risager, ``MHV-vertices for gravity
amplitudes,'' JHEP {\bf 0601} (2006) 009 [arXiv:hep-th/0509016];
P.~Benincasa, C.~Boucher-Veronneau and F.~Cachazo, ``Taming tree
amplitudes in general relativity,'' JHEP {\bf 0711} (2007) 057
[arXiv:hep-th/0702032]; N.~Arkani-Hamed and J.~Kaplan, ``On Tree
Amplitudes in Gauge Theory and Gravity,'' JHEP {\bf 0804} (2008) 076
[arXiv:0801.2385 [hep-th]].


%\cite{BCFW:loop-level}
\bibitem{BCFW:loop-level}
 Z.~Bern, L.~J.~Dixon and D.~A.~Kosower,
  ``On-shell recurrence relations for one-loop QCD amplitudes,''
  Phys.\ Rev.\  D {\bf 71} (2005) 105013;
  N.~Arkani-Hamed, J.~L.~Bourjaily, F.~Cachazo, S.~Caron-Huot and J.~Trnka,
  ``The All-Loop Integrand For Scattering Amplitudes in Planar N=4 SYM,''
  arXiv:1008.2958 [hep-th];
 R.~H.~Boels,
  ``On BCFW shifts of integrands and integrals,''
  arXiv:1008.3101 [hep-th].


%\cite{BCFW:string}
\bibitem{BCFW:string}
R.~Boels, K.~J.~Larsen, N.~A.~Obers and M.~Vonk,
  ``MHV, CSW and BCFW: field theory structures in string theory amplitudes,''
  JHEP {\bf 0811} (2008) 015
  [arXiv:0808.2598 [hep-th]].
  C.~Cheung, D.~O'Connell and B.~Wecht,
  ``BCFW Recursion Relations and String Theory,''
  JHEP {\bf 1009} (2010) 052
  [arXiv:1002.4674 [hep-th]].
R.~H.~Boels, D.~Marmiroli and N.~A.~Obers,
  ``On-shell Recursion in String Theory,''
  arXiv:1002.5029 [hep-th].
A.~Fotopoulos and N.~Prezas,
  ``Pomerons and BCFW recursion relations for strings on D-branes,''
  arXiv:1009.3903 [hep-th].
A.~Fotopoulos,
  ``BCFW construction of the Veneziano Amplitude,''
  arXiv:1010.6265 [hep-th].




%\cite{BCFW:boundary}
\bibitem{BCFW:boundary}
B.~Feng, J.~Wang, Y.~Wang and Z.~Zhang,
  ``BCFW Recursion Relation with Nonzero Boundary Contribution,''
  JHEP {\bf 1001} (2010) 019
  [arXiv:0911.0301 [hep-th]];
B.~Feng and C.~Y.~Liu,
  ``A note on the boundary contribution with bad deformation in gauge theory,''
  JHEP {\bf 1007} (2010) 093
  [arXiv:1004.1282 [hep-th]].

%\cite{S-matrix-program}
\bibitem{S-matrix-program}
P.~Benincasa and F.~Cachazo,
  ``Consistency Conditions on the S-Matrix of Massless Particles,''
  arXiv:0705.4305 [hep-th];
N.~Arkani-Hamed, F.~Cachazo and J.~Kaplan,
  ``What is the Simplest Quantum Field Theory?,''
  JHEP {\bf 1009} (2010) 016
  [arXiv:0808.1446 [hep-th]].











%
%\cite{Chen:2009tr}
\bibitem{Chen:2009tr}
  Y.~X.~Chen, Y.~J.~Du and Q.~Ma,
  ``Relations Between Closed String Amplitudes at Higher-order Tree Level and
  Open String Amplitudes,''
  Nucl.\ Phys.\  B {\bf 824} (2010) 314
  [arXiv:0901.1163 [hep-th]].
  %%CITATION = NUPHA,B824,314;%%


%\cite{Chen:2010sr}
\bibitem{Chen:2010sr}
  Y.~X.~Chen, Y.~J.~Du and Q.~Ma,
  ``Disk relations for tree amplitudes in minimal coupling theory of gauge
  field and gravity,''
  Nucl.\ Phys.\  B {\bf 833} (2010) 28
  [arXiv:1001.0060 [hep-th]].
  %%CITATION = NUPHA,B833,28;%%


\end{thebibliography}
\end{document}